\documentclass[aps,twocolumn,pra,tightenlines,floatfix,showpacs]{revtex4}

\usepackage[dvips]{graphicx}
\usepackage[english]{babel}
\usepackage{amsmath}
\usepackage{amssymb}
\usepackage{times}

\newcommand{\bs}{\begin{split}}
\newcommand{\es}{\end{split}}
\newcommand{\be}{\begin{equation}}
\newcommand{\ee}{\end{equation}}
\newcommand{\ba}{\begin{eqnarray}}
\newcommand{\ea}{\end{eqnarray}}

\begin{document}

\title{Fermions with attractive interactions on optical lattices and
implications for correlated systems}

\author{Chih-Chun Chien, Qijin Chen, and K.  Levin}

\affiliation{James Franck Institute and Department of Physics,
 University of Chicago, Chicago, Illinois 60637}

\date{\today}

\begin{abstract} 
In this paper we address the behavior of the superfluid transition
  temperature $T_c$ in the attractive Hubbard model.
The attractive Hubbard model can be
regarded as the generalization of BCS to Bose Einstein condensation
(BEC) crossover to a lattice and as such, may have implications for
future optical lattice studies. Nevertheless, the BEC limit of the Hubbard model is
very different from that of the more well studied
Fermi gases, owing to the
strong inter-site repulsion between pairs.
Here we address systematically the effects of pairing fluctuations for
all filling fractions over the entire range of attractive
interaction strength. 
A central
conclusion of our work is that in a lattice, around half filling, the
smooth evolution from the BCS to the BEC limits is interrupted:
$T_c$ vanishes when the system
approaches the bosonic regime with increasing interaction strength. We
suggest that this interruption of crossover may
signal a quantum critical transition to another form of superfluid not
continuously connected to a BCS-like phase.
A simple variational ansatz for an alternate
ground state in this more strongly coupled superfluid is presented.  A generalization
of
the ($s$-wave) Hubbard model to $d$-wave
pairing allows us to address issues of relevance to high $T_c$
superconductivity. The
phase diagram
(representing the pairing or pseudogap onset temperature $T^*$ and $T_c$) shows that here too, one observes a
vanishing of $T_c$ when $T^*$ becomes sufficiently large.  Given this
predicted breakdown of the crossover, and given the striking
similarity to features of the cuprate phase diagram, we suggest that
future experiments on ultracold fermions in optical lattices should
not be exclusively limited to the repulsive Hubbard model, but should
address the attractive model in order to elucidate features of high
temperature superconductivity.
\end{abstract}

\pacs{03.75.Hh, 03.75.Ss, 74.20.-z }

\maketitle

---------------------------------------------------------------------- 
\section{Introduction} 
Recent experiments on ultracold Fermi gases of $^{40}$K \cite{Stoferle}
and $^6$Li \cite{MITlattice} in the presence of an optical lattice are
creating considerable interest in the community.  These experiments can
be viewed as quantum simulations \cite{Demler2,GeorgesReview} of
important condensed matter problems such as the repulsive and attractive
fermion Hubbard models. In these experiments one can vary (albeit, not
independently) both the on-site interaction strength between fermions
and the lattice depth and in this way change the interaction $U$ and
hopping $t$.  In the case of \textit{attractive} interaction $U$, one
can view the Hubbard model as generalizing the important problem of
BCS-- Bose-Einstein condensation (BEC) crossover
\cite{Leggett,NSR,Randeriareview} to the case of a lattice.  This
crossover study represents an exciting research area because it provides
a means of extending what is arguably the paradigm of all theories in
condensed matter physics (BCS theory) to a far more general situation.

Many have argued that \cite{LeeReview} the key to
high temperature superconductivity comes from the \textit{repulsive}
Hubbard model and there is a growing impetus to use cold atoms to
address the issue of whether this repulsive model can ultimately lead to
$d$ wave superconductivity \cite{Pines1997,Scalapino86}.  Nevertheless, our paper is based on the
premise that we have potentially as much to learn about high $T_c$
superconductors from the attractive as from the repulsive Hubbard case.
This attractive model represents a generalization of BCS-BEC crossover
which is of current interest in atomic Fermi systems
\cite{ourreview,StringariReview} to now include the presence of an
optical lattice.  We will, in addition to the $s$-wave case, consider a
natural generalization to the $d$-wave pairing symmetry.  The former,
however, has been more extensively addressed in the literature and is of
considerable value since the experimental counterpart is more
straightforward. The latter is more relevant to high $T_c$ superconductors.

We note that the BCS-BEC crossover scenario is argued
\cite{LeggettNature} to be relevant to high temperature superconductors
because of their anomalously short coherence length.  Additional support
for the potential relevance of the attractive Hubbard model (AHM) to the
copper oxides comes from the fact that this crossover scenario naturally
leads to a ``pseudogap'' phase which is quite prominent in the cuprates,
although there is still no consensus on its origin.
\cite{ourreview,Varenna,LeeReview} The pseudogap state corresponds to a
regime (above $T_c$) where there is a gap for exciting fermionic
quasi-particles.  Within the crossover scenario a natural interpretation
of this gap associates it with the existence of metastable (sometimes
called pre-formed) pairs arising from the stronger-than-BCS attractive
interaction. These noncondensed pairs require an added energy in order
to break them apart and create fermionic excitations.

A central goal of this paper is to present a phase diagram for the
various regions of $d$- and $s$-wave superfluid stability as a function
of interaction strength and band filling. In this context, we also
present a phase diagram indicating how the characteristic temperatures,
i.e., the pseudogap onset temperature $T^*$ and the condensation
temperature $T_c$, vary with the strength of the attractive interaction
in the $d$-wave case.  When the attraction gets progressively stronger
(as measured by the size of $T^*$), $T_c$ begins to decrease and
ultimately vanishes.  This behavior is strikingly similar to that found
in the cuprate phase diagram (\cite{LeeReview}, especially Fig.~7)
where, via larger $T^*$, one can associate stronger attraction with
reduced doping concentration.  These observations suggest
that future cold gas experiments should not focus solely on the
repulsive Hubbard model, but also give extensive attention to the
attractive case.

In so far as they relate to the high temperature superconductors, the
attractive and repulsive Hubbard models should be directly compared.  In
particular, these two models may be essentially equivalent if in the
attractive case, the interaction is presumed to be $d$-wave.  The
repulsive Hubbard model is thought to give rise to a $d$-wave pairing
\cite{Scalapino86,Pines1997,LeeReview} although this has yet to be
conclusively demonstrated.  Given that the pairing symmetry in the
cuprates is known to be $d$-wave and that the pair size is anomalously
small, it is certainly of equal interest to study a $d$-wave
generalization of the \textit{attractive} Hubbard model, as we do
here. Within this latter context no assumption is made about the
microscopic origin of the attractive interaction.  Quite possibly,
however, this inter-site attraction comes the Coulomb interaction which
is repulsive at short distances, but can be even more complex than
contemplated \cite{Liu,Leggett2} by the simple repulsive Hubbard model.

While there have been a number of numerical studies of the $s$-wave
pairing model \cite{KellerTmatrix,Beck02,Scalettar2D}, along with
approaches based on dynamical mean field theory
\cite{KellerDMFT,CaponeDMFT1,CaponeDMFT2,WernerPRL}, ours is the only
systematic study over all filling fractions and over the entire range
of attractive interactions. We investigate the stability of the
BCS-Leggett like ground state (applied to the lattice).  This state is
a natural one to consider for the purposes of ultimately shedding
light on the cuprates, since there is a strong belief in the community
\cite{ourreview,Varenna,LeeReview} that when superconductivity is
present, it is in many aspects not so different from the $d$-wave BCS
state. In this way we argue that, while more sophisticated ground state
wave functions might be of great interest for reaching a general
understanding, they may not be of specific relevance to the cuprates.

We find that, even though the strict $T = 0$ mean-field
equations suggest that superfluidity can occur everywhere, fluctuation
effects lead to a vanishing $T_c$ over an extended range sufficiently
near half filling and for moderately strong attraction.  An important
clue underlying this breakdown of BCS-BEC crossover is that it occurs
very close to the point where the fermionic chemical potential $\mu $
changes sign. We naturally interpret the $\mu =0$ point as corresponding
to the transition into a regime where the system is effectively bosonic.
Here the ``bosons'' are associated with condensed or noncondensed
(finite momentum) Cooper pairs.

To gain further insight, we are motivated, then, to consider an
\emph{effective} bosonic model representing the AHM.  Indeed, when one expands this Hamiltonian in terms of $t^2/|U|$
the model which emerges contains bosonic hopping as well as an
inter-site repulsion of precisely the same magnitude. At low density the
repulsion is not important, and the system is expected to support
BCS-like superfluidity with a $T_c$ of order $t^2/|U|$, as originally
conjectured in the important paper by Nozieres and Schmitt-Rink (NSR)
\cite{NSR}. However, closer to half filling 
in the $t^2/U$-expanded bosonic Hamiltonian the inter-boson interactions
are \textit{strong}.  We emphasize that in this simple BCS-Leggett
ground state the interaction between bosons is implicitly taken to be
weak. More sophisticated wave functions are required to capture the
effects of stronger inter-boson interactions \cite{Shina2} which must go
even beyond Bogoliubov level theory to be consistent with the physics of
the $t^2/U$-expanded Hubbard model.  This, then, provides an explanation
for the failure of this simple BCS trial wave function to support a
superfluid phase in the BEC regime on a lattice.  Finally, these studies
are extended to the $d$-wave case as well, where the breakdown of the
BCS-Leggett like superfluid occurs well before the chemical potential
becomes negative.

There has been a substantial literature on BCS-BEC crossover on the
subject of Fermi gases \cite{ourreview,Varenna,StringariReview} with tunable
attractive interactions.  There is also considerable interest in whether
an analogous crossover can be observed for fermions on a lattice
\cite{Chienlattice1,Stooflattice}.  To effect this crossover we note
that there are two ways of increasing the dimensionless attraction
$U/t$, either by changing the lattice depth or changing the on-site
attraction $U$ by exploiting a Feshbach resonance.  For the present
purposes of addressing BCS-BEC crossover in the context of the
AHM \cite{Fedichev,GeorgesReview} (which we have argued
above may contain information about high temperature superconductors) we
should not exploit these Feshbach effects. 
Rather, the hopping $t$ and the on-site attraction $U$ are varied by changing
parameters of the optical lattice.  Thus, our conclusions about the
interruption of BCS-BEC crossover near the onset of BEC are entirely
associated with the Hubbard model and do not pertain if one uses a
Feshbach resonance to create bosonic degrees of freedom, prior to
applying an optical lattice. In this alternative circumstance the more
appropriate many body model is the Bose Hubbard model which we do not
consider here.

There is an extensive literature which addresses the AHM, (\cite{MicnasRMP} and references therein), principally in
the two-dimensional case because of associations with high-temperature
superconductivity. In a related fashion, the possibility of observing
(bosonic-like) Mott insulating states at full filling 
has been the motivation for studies from a number of groups
\cite{Carr05,Zhou05,Holattice,Nikolic}, particularly from the atomic
physics community.  There has also been a focus on charge density wave
states which are energetically degenerate with the superfluid phase at
precisely half filling and may compete or co-exist with superfluidity
\cite{MicnasRMP,Burkov} slightly away.  Indeed, a supersolid phase
contemplated in this literature is viewed as a mixture of charge density
wave and superfluidity.  Finally, we note that there is also a
methodology for addressing the AHM via a mapping to the repulsive
Hubbard case. Except at exactly 1/2 filling, the counterpart repulsive
model must be solved in the presence of a rather complicated constraint,
which is difficult to implement. A general theorem which states that, in
the AHM, the ground state contains no magnetic
order \cite{Lieb}, must be imposed \cite{MicnasRMP} in any study based
on this mapping.

For the most part these previous studies have been at zero
temperature. Here we focus on general temperatures, $T$.  By choosing to
address the superfluid transition temperature, $T_c$, we effectively
introduce fluctuation contributions in a fashion that is 
consistent with the BCS-Leggett ground state and capable of addressing
finite temperatures.  Without these fluctuations, in the $s$-wave case,
it appears that a BCS-Leggett superfluid ground state is stable for all
parameters.  By considering the entire range of (attractive) interaction
strengths and filling factors, we enter into regimes which have also not
been addressed by complementary numerical or mean field techniques.

Our theoretical scheme is based on a particular $T$-matrix approximation
for the pairing fluctuations which is compatible with this simplest
ground state. Alternative $T$-matrix schemes have been applied to
attractive Hubbard models in a number of different variations.  The
transition temperature $T_c$ was estimated in \cite{KellerTmatrix}
within an approach in which there is no pseudogap phase in the normal
state.  By contrast, here as in earlier work
\cite{Chen1,Micnaslattice,Chienlattice1}, we use a scheme in which
superconductivity emerges in the presence of a pseudogap. 
Alternative approaches based on dynamical mean field theory
\cite{KellerDMFT,CaponeDMFT1,CaponeDMFT2} in three dimensions (3D) 
and on quantum Monte Carlo simulations in 3D \cite{Beck02} or in 2D \cite{Scalettar2D} have also addressed
the size of the transition temperature.  For intermediate attraction
strengths, a maximum in the $T_c$ curves is found, which some have
argued \cite{Fedichev,Stooflattice} corresponds to the regime where
BCS-BEC crossover occurs. However, we stress that this maximum appears
deep in the fermionic regime, quite far from where the fermionic chemical
potential $\mu$ becomes negative.

It should be noted that there are also extensive studies of the Bose
Hubbard model \cite{MPFisher,Rokhsar,Scalettar} which are viewed as
relevant to the strongly attractive regimes of the fermion AHM.  We
emphasize here, however, that there is an important distinction between the
composite boson BEC limit and the Bose Hubbard model, due to the
different commutation properties of the field operators for fermion
pairs versus those for true bosons. This difference is less important in
the very low filling regime, but it cannot in general be ignored.
Finally, on the subject of BCS-BEC crossover for the $d$-wave case, there
are earlier studies \cite{Chen1,Hertog} in the literature.  In
Ref.~\cite{Hertog} a simplified model addressed the 2D square lattice at
$T=0$.

\section{General Formalism: BCS-BEC Crossover on a Lattice}

The AHM Hamiltonian is given by 
\begin{equation} 
  H^{\mbox{AHM}}=\sum_{<i,j>,\sigma}t_{ij}c^{\dagger}_{i\sigma}c_{j\sigma}+\frac{1}{2}U\sum_{i}(n_i-1)^2. 
\end{equation} 
Here $\langle i,j\rangle$ denotes nearest neighbors, $t_{ij}$ denotes
the hopping coefficient, $\sigma=\uparrow, \downarrow$ denotes spins,
$U<0$ is the on-site attractive coupling constant,
$n_i=n_{i\uparrow}+n_{i\downarrow}$, and
$n_{i\sigma}=c^{\dagger}_{i\sigma}c_{i\sigma}$. We consider the case
where there is an equal population of both fermion spin states.  The
fermion filling factor is $n=\sum_i\langle n_i\rangle/N$, where $N$ is
the total number of sites on the lattice. In the single-band
tight-binding approximation only the case $0<n<1$ needs to be considered
due to particle-hole symmetry.

In order to include $d$-wave pairing as well,
it is convenient to consider a generalization of the AHM in a momentum
space representation.
\begin{eqnarray} 
& &H-\mu\hat{N} = \sum_{{\bf k}\sigma} \xi_{\bf 
k} c^{\dag}_{{\bf k}\sigma} c^{\ }_{{\bf k}\sigma} \nonumber \\ & & + 
\sum_{\bf k k' q} V_{\bf k, k'} c^{\dag}_{{\bf k}+{\bf q}/2\uparrow} 
c^{\dag}_{-{\bf k}+{\bf q}/2\downarrow} c^{\ }_{-{\bf k'}+{\bf 
q}/2\downarrow} c^{\ }_{{\bf k'}+{\bf q}/2\uparrow}. \label{eq:Hk} 
\end{eqnarray} 
Here $\xi_{\mathbf{k}}=\epsilon_{\mathbf{k}}-\mu$, $\mu$ is the fermion
chemical potential, $\hat{N}=\sum_{\bf k}c^{\dag}_{{\bf k}}c^{\ }_{{\bf
    k}}$ is the total number operator, $\epsilon_{\mathbf{k}}$ is the
energy in the tight-binding band.  The interaction in momentum space
assumes the separable form
$V_{\mathbf{k},\mathbf{k}^{\prime}}=U_{\alpha}\varphi_{\mathbf{k}}\varphi_{\mathbf{k}^{\prime}}$. Here
$U_{\alpha}=U$ and $\varphi_{\mathbf{k}}=1$ for $s$-wave and
$U_{\alpha}=U_{d}$ and $\varphi_{\mathbf{k}}=\cos k_x-\cos k_y$ for
$d$-wave pairing. 
For the $s$-wave case, we consider isotropic 3D square lattices with
$\epsilon_{\mathbf{k}}=2t(3-\cos k_x-\cos k_y -\cos k_z)$, where $t$,
the hopping integral, serves as the unit of energy.  For $d$-wave
pairing we consider quasi-2D square lattices with
$\epsilon_{\mathbf{k}}=2t_{\parallel}(2-\cos k_x-\cos
k_y)+2t_{\perp}(1-\cos k_z)$, where $t_{\parallel}$ and $t_{\perp}$ are
the in-plane and out-of-plane hopping integrals, respectively.  We use
$t_{\parallel}$ as the unit of energy, and presume, for definiteness, an
anisotropic ratio $t_{\perp}/t_{\parallel}=0.01$, which is reasonable
for the cuprate superconductors \cite{Chen1}.

When the attraction is weak ($|U_{\alpha}|/t\rightarrow 0$), the
superfluid phase consists of loosely bound pairs, as in a BCS model. As
$|U_{\alpha}|$ is progressively increased the pairs become more tightly
bound and the system crosses over over to a BEC-based description in which
there are composite bosons confined to a lattice. 
The simplest possible ground state for describing this crossover on a
lattice is that associated with BCS-Leggett theory \cite{Leggett}. 
We will adopt this ground state here:
\begin{equation}
\label{eq:kSF}
|BCS\rangle=\prod_{\mathbf{k}}(u_{\mathbf{k}}+v_{\mathbf{k}}c^{\dagger}_{\mathbf{k}\uparrow}c_{-\mathbf{k}\downarrow})|0\rangle.
\end{equation} 
Here the coefficients $u^{2}_{\mathbf{k}}, v^{2}_{\mathbf{k}}=(1\pm
\xi_{\mathbf{k}}/E_{\mathbf{k}})/2$ and the quasiparticle dispersion 
$E_{\mathbf{k}}=\sqrt{\xi_{\mathbf{k}}^2+\Delta^{2}\varphi_{\mathbf{k}}^2}$
. This ground state presumes that the
condensation is complete and is, more generally, appropriate only for
weakly interacting bosons or Cooper pairs. In the single trap
experiments the BEC asymptote is indeed associated with free bosons and
one might expect this wave function to be a reasonable starting point.
On the lattice, in the BEC limit, however, the repulsive interaction is equal to the
effective kinetic energy of the pairs (as will be discussed below). Therefore, in a
properly self-consistent theory, one might expect to see a break down of
this ansatz for the ground state wave function near half filling where the effects of 
inter-pair repulsion are no longer negligible.

\subsection{Present T-matrix Scheme}

To solve for $T_c$ in a fashion consistent with the known ground state
constraints \cite{Leggett,NSR} of the gap and fermion number equations,
we follow earlier calculations \cite{ourreview,Varenna,Chienlattice1},
based on a $T$-matrix theory for pairing fluctuations. The $T$-matrix is
given by
$t(K,K^{\prime},Q)=t(Q)\varphi_{\mathbf{k}}\varphi_{\mathbf{k}^{\prime}}$,
and contains two contributions, $t(Q)=t_{sc}(Q)+t_{pg}(Q)$, describing
the condensed and noncondensed pairs, respectively. Here we take $K\equiv
(i\omega_n, \mathbf{k})$, $K^{\prime}\equiv
(i\omega_{n^{\prime}}, \mathbf{k}^{\prime})$, and $Q\equiv (i\Omega_m,\mathbf{q})$ as
four-vectors. We use $\omega_n$ and $\Omega_m$ to represent the
Matsubara frequencies for fermions and bosons, respectively, and
$\sum_{K}\equiv T\sum_{\omega_n}\sum_{\mathbf{k}}$.
It can be shown that if one takes
$t_{sc}(Q)=-(\Delta_{sc}^2/T)\delta(Q)$, as $T\rightarrow 0$ the
$T$-matrix theory is consistent with the constraints in the BCS-Leggett
ground state. Here $\Delta_{sc}$ is the order parameter which vanishes at
$T_c$ but is equal to the total gap at $T=0$. The contribution from the
noncondensed pairs, which contains the sum of particle-particle
scattering ladder diagrams, is
\begin{equation}
t_{pg}(Q)= \frac{U_{\alpha}}{1+U_{\alpha}\chi(Q)}, 
\end{equation}
and the pair susceptibility is
$\chi(Q)=\sum_{\mathbf{k}}G(K)G_0(Q-K)\varphi_{\mathbf{k}-\mathbf{q}/2}^2$.
Here $G_{0}(K)=(i\omega_n -\xi_{\mathbf{k}})^{-1}$ is the bare Green's
function and $G(K)$ is its dressed counterpart.

To define the appropriate dressed Greens functions, we adopt the usual
$T$-matrix expression for the fermion self-energy given by
$\Sigma(K)=G_{0}^{-1}(K)-G^{-1}(K)=\sum_{Q}t(Q)G_{0}(Q-K)\varphi^{2}_{\mathbf{k}-\mathbf{q}/2}$. 
This self-energy can be approximated as
$\Sigma(K)=-\Delta^{2}G_{0}(-K)\varphi_{\mathbf{k}}^2$, where
$\Delta^2=\Delta_{sc}^2+\Delta_{pg}^2$ and the pseudogap contribution is
\begin{equation}\label{eq:Dpg}
\Delta_{pg}^{2} \equiv -\sum_{Q}t_{pg}(Q) 
\end{equation}
The total fermion number equation is given by $n=2\sum_{K}G(K)$, or
\begin{equation} 
\label{eq:neq} 
n = 
\sum_{\mathbf{k}}
\left[\left(1-\frac{\xi_{\mathbf{k}}}{E_{\mathbf{k}}}\right) +
  2f(E_{\mathbf{k}})\left(\frac{\xi_{\mathbf{k}}}{E_{\mathbf{k}}}\right)\right]. 
\end{equation}

The BEC condition requires that the pairs have zero chemical potential
at and below $T_c$: $t_{pg}^{-1}(0)=U_{\alpha}^{-1}+\chi(0)=0$, 
i.e., 
\begin{equation}\label{eq:geq}
  t_{pg}^{-1}(0)=U_{\alpha}^{-1}+\sum_{\mathbf{k}}\frac{1-2f(E_{\mathbf{k}})}{2E_{\mathbf{k}}}\varphi_{\mathbf{k}}^{2}=0,
  \qquad  T \leq T_c.
\end{equation}
For $T\le T_c$, to satisfy this BCS-like gap equation (which naturally
emerges for $\Delta$ within the present $T$-matrix scheme), $\Delta$
must necessarily contain a contribution from both a non-zero superfluid
order parameter, $\Delta_{sc}$, and a contribution associated with
noncondensed pairs, $\Delta_{pg}$.  In this way and in the weak coupling
limit, this approach represents a re-interpretation of BCS theory which
underlines the strong similarity to the treatment of ideal gas BEC. The
contribution from noncondensed pairs enters through a gap equation, not
a number equation, however. The transition temperature $T_c$ is the
temperature above which $\Delta_{sc}$ vanishes and is determined by
solving Eqs.~(\ref{eq:Dpg}), (\ref{eq:neq}), and (\ref{eq:geq})
self-consistently.

Below $T_c$ the $T$-matrix may be expanded as $t_{pg}^{-1}\approx
a_1\Omega^2+a_0\Omega-\xi^2q^2 + i\Gamma_{\mathbf{q}}$ after analytic
continuation.  This simplifies the evaluation of Eq.~(\ref{eq:Dpg}).
The effect of the $a_1\Omega^2$ term can be neglected except in a narrow
regime near half filling, where particle-hole symmetry leads to
$a_0\rightarrow 0$. The contribution of pairs is dominated by those near
the bottom of their energy band so the dispersion is further
approximated as $\Omega_{\mathbf{q}}=\xi^2q^2/a_0 = q^2/2M_p$, where $M_p$
is the effective mass of the pairs on a lattice. Note that
$\Gamma_{\mathbf{q}} \ll \xi^2q^2$ in the long wavelength limit so that
it is set to 0 in our numerical calculations.

\subsection{The strong attraction limit}

For the case of $s$-wave pairing, it is useful at this stage to study an
approximated Hamiltonian in the limit that $|U|/t$ is finite but very
large so that the hopping term can be treated as a
perturbation. Following Refs.~\cite{Rob,S64} and dropping overall
constants, the Hamiltonian can be rewritten as
\begin{equation} 
H_{eff}=-\sum_{<i,j>}Jb^{\dagger}_{i}b_{j}+\sum_{<i,j>}Jn_{b_i}n_{b_j}.
\label{eq:Heff}
\end{equation} 
Here $J=2t^2/|U|$, $b_i=c_{i\downarrow}c_{i\uparrow}$,
$b^{\dagger}_i=c^{\dagger}_{i\uparrow}c^{\dagger}_{i\downarrow}$, and
$n_{b_i}=b^{\dagger}_i b_i$. The pair operators $b_i$ and
$b^{\dagger}_i$ are not strictly boson annihilation and creation
operators, since their commutator $[b_i,b^{\dagger}_i]=1-n_i$, and
$\{b_i, b_i\}=0$, where
$n_i$ represents the number of fermions at site $i$.
The Pauli
principle insures that these ``bosons'' are hard core bosons. 
This effective Hamiltonian is equivalent to an XXZ magnetic model 
with an effective external field in the $z$ direction; here
the average magnetization must have a fixed value equal to $(n-1)/2$ (see \cite{MicnasRMP}).

It is important to stress that Eq.~(\ref{eq:Heff}) contains a nearest
neighbor inter-site repulsion which is of the same value $J$ 
as the boson hopping term. This inter-site repulsion becomes
progressively more important when the fermion filling is close to one
half.  While the fact that the kinetic energy in Eq.~(\ref{eq:Heff})
varies as $t^2/U$ is relatively straightforward to understand, the
origin of the inter-site repulsion is more subtle.  This term arises
\cite{NSR} from the energy lowering associated with virtual hopping of
fermions. 
Clearly the Pauli principle leads to a constrained hopping; if a pair
has an occupied nearest neighbor site, then hopping will be suppressed,
thereby raising the energy.  In this way we see that there is an
effective inter-site repulsion between the pairs.

The Hamiltonian of Eq.~(\ref{eq:Heff}) should be contrasted with the
boson Hubbard model (BHM) which has been widely studied \cite{MPFisher}
in the context of the Mott insulator-superfluid transition. The BHM
corresponds to true bosons on a lattice, where the kinetic energy
contribution and the \textit{on-site} repulsion $U_{B}$ can be varied
independently:
\begin{equation} 
  H^{BHM}=-\sum_{<i,j>}J_{B}\bar{b}^{\dagger}_{i}\bar{b}_{j}+\sum_{i}U_{B}n_{\bar{b}_i}n_{\bar{b}_i}. 
\end{equation}

\section{Numerical Results for the $s$-wave case}

The system of equations (\ref{eq:Dpg})-(\ref{eq:geq}) can be readily
solved to yield the transition temperature $T_c$ for the Leggett-BCS
state and with variable $t/U$ and filling factors.  We, again, stress
that there is an important difference \cite{NSR} between the
BEC limit in a gas and on a lattice. For the latter the transition
temperature becomes zero at arbitrarily large attractive interactions.
This is associated with the fact that the effective mass of the pair is
infinite; pairs cannot hop without an intermediate unbinding which
becomes prohibitively costly at large attraction.  Very early on
\cite{NSR} it was anticipated that $T_c$ would vary as
$t^2/|U|$ in the deep BEC.
We emphasize here that the the validity of the scaling, $T_c \propto
t^2/|U|$, requires minimally that the fermionic chemical potential $\mu$
be negative, i.e., the system must be in the bosonic regime.  Indeed, 
as in earlier work \cite{Chen1} we
find this dependence for $T_c$ (but only) in the low density limit.

\begin{figure} 
\centerline{\includegraphics[clip,width=3.4in]{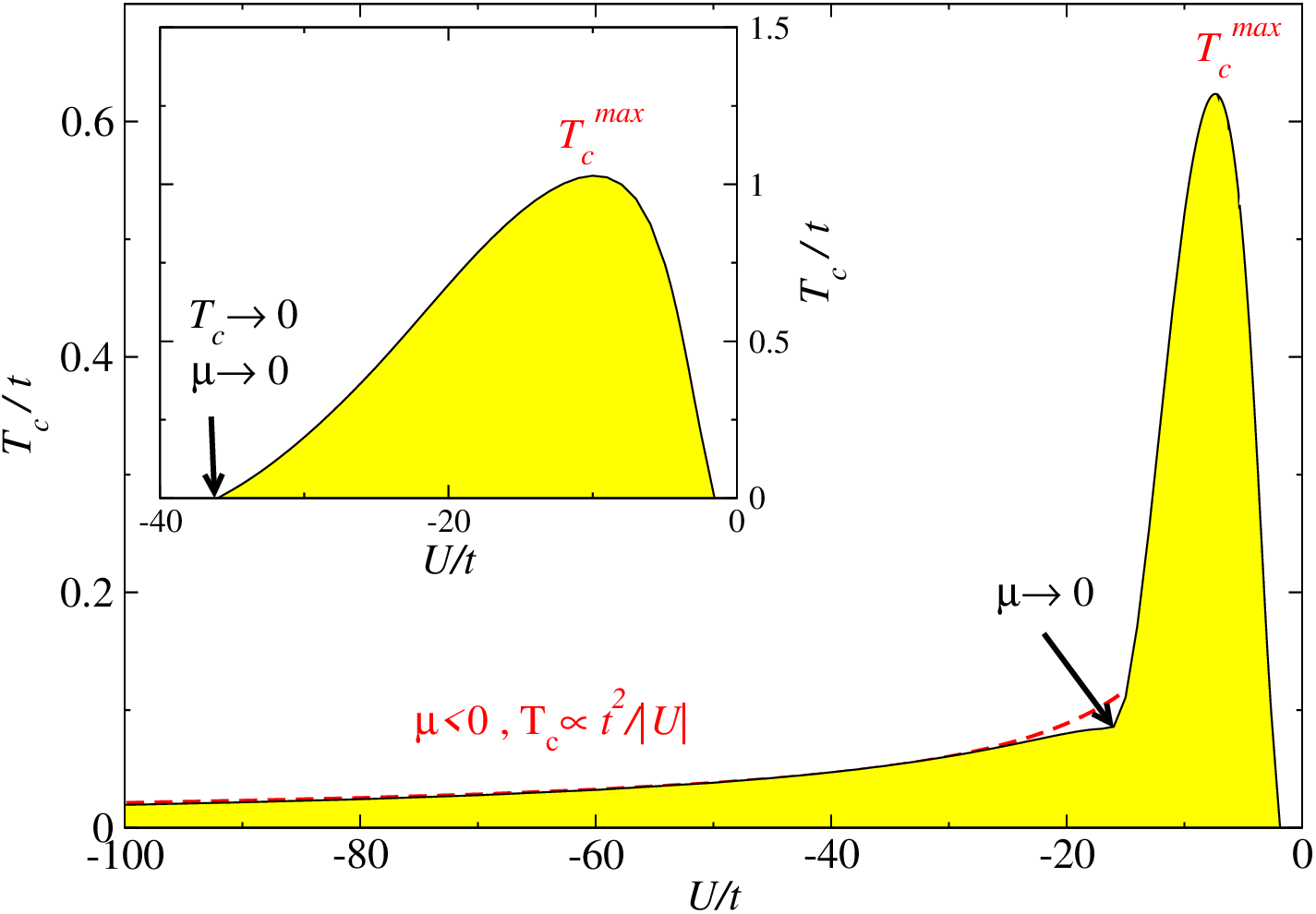}} 
\caption{(Color online) $T_c$ curve for $s$-wave pairing 
on a 3D lattice for $n=0.3$ (black solid line). The tail is fitted to 
the functional form $t^2/|U|$ (red dash line). Inset: $T_c$ curve for 
$n=0.7$.} 
\label{fig:L1n03}
\end{figure}

Figure~\ref{fig:L1n03} presents plots of $T_c$ as a function of $U/t$ at
$n=0.3$ and $n=0.7$ (inset). For the first case, at low filling, $T_c$
has a maximum in the regime where $\mu>0$; once $\mu$ becomes negative
we find a long tail with the expected $t^2/|U|$ dependence. This general
behavior is
consistent with earlier work \cite{Chen1,Chienlattice1,Micnaslattice}.
It should be noted that, using rather different formalisms
from that discussed here, others \cite{KellerDMFT,WernerPRL} have reported
this asymptotic $t^2/U$ behavior, but have never correlated it with the
bosonic regime, in the sense of requiring a negative $\mu$.
The plot in the inset emphasizes this point.  At high $n$ this figure
shows that $T_c $ vanishes right before the system crosses over into the
bosonic regime.  Also labeled on the plots for both cases is
$T_c^{max}$, where the transition temperature reaches a maximum.  As 
noted above, this maximum occurs at intermediate values of $|U|/t$
for all filling. As for the unitary limit in the case of homogeneous Fermi gases
\cite{ourreview,Varenna}, it occurs deep within the fermionic regime,
where the chemical potential is still positive.

\begin{figure} 
\centerline{\includegraphics[clip,width=3.4in]{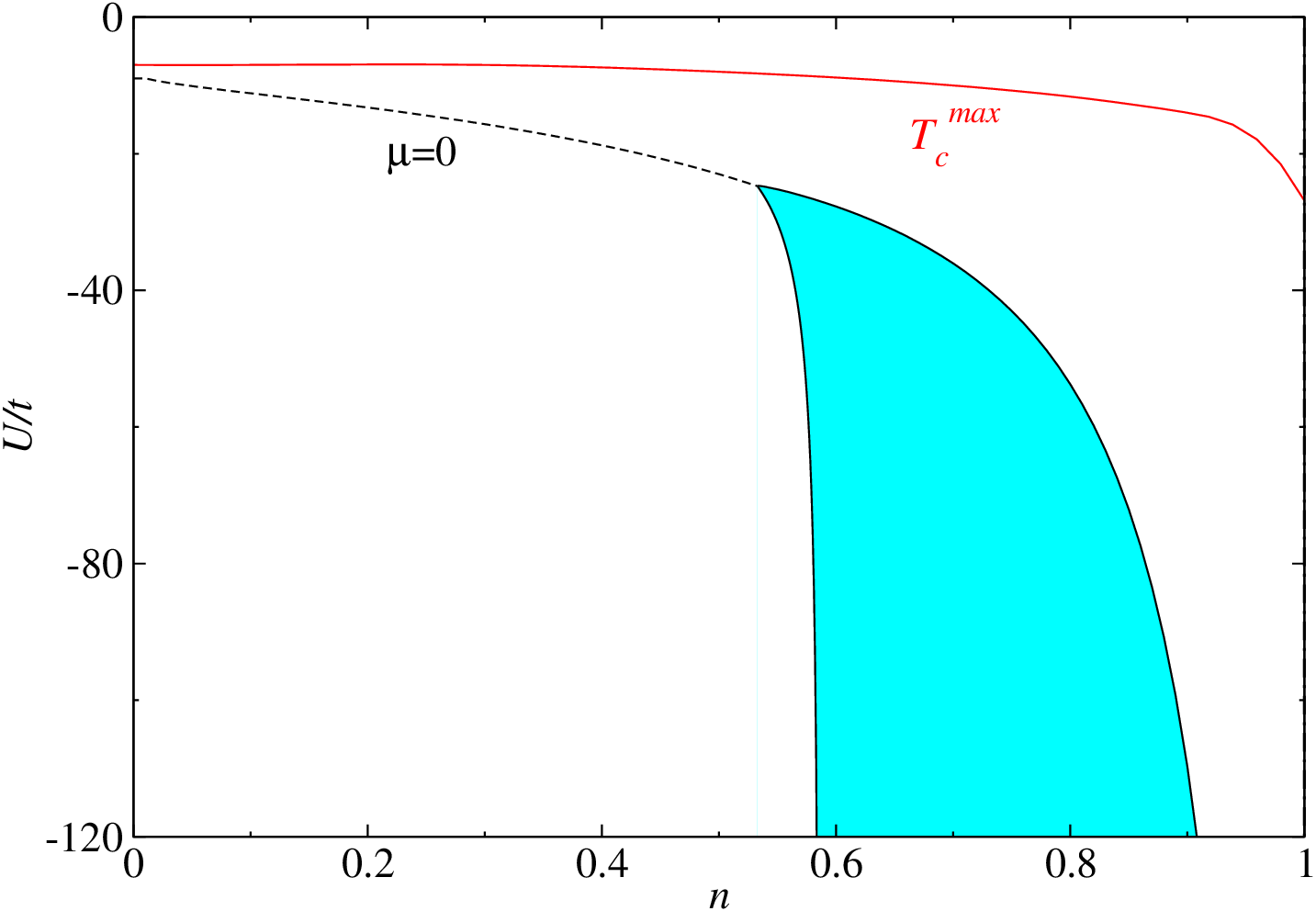}} 
\caption{(Color online) $T\rightarrow 0$ phase diagram associated with
  the BCS-Leggett state for the AHM.  The shaded
  region shows where superfluidity breaks down due to strong pairing
  fluctuations. The red line shows the trace of $T_c^{max}$ in $T_c$
  plots as shown in Fig.~\ref{fig:L1n03}. The black dashed line shows
  where $\mu=0$,
  corresponding to the onset of a $t^2/|U|$ dependence in $T_c$,
  associated with low filling. At 
  $n=1$, there is an exact particle-hole symmetry.}
\label{fig:L1T0}
\end{figure}

Figure ~\ref{fig:L1T0} summarizes our phase diagram in the $U/t-n$ plane
for the AHM with the shaded regimes indicating
where the calculated $T_c$ vanishes. Also indicated is the location of
$T_c^{max}$ as shown in Fig.~\ref{fig:L1n03} as a function of $n$. 
It is relatively constant in $n$, as observed by other groups
\cite{KellerTmatrix,KellerDMFT,Beck02}.  
Because of
particle-hole symmetry at $n=1$, here we only need to focus on the $n<1$
half of the phase diagram.

We note that the upper boundary
of the shaded region marking the breakdown of this superfluid phase is
consistently near the $\mu=0$ line (below which Cooper pairs start to
behave like composite bosons), as might have been expected from the
previous figure. 
Interestingly enough, this upper boundary (say, for
$n \approx 0.5$) is not so far from the predicted values for $U/t= 35$
at which the superfluid-Mott insulator transition takes place for the BHM \cite{BlochRMP}. Note that for the BHM, $U$ must be positive in order to
stabilize Mott phases which derive from
strong on-site inter-boson repulsion. Finally, we note that particle-hole
symmetry near half filling effectively pins the chemical potential near
the band center and the system has difficulty reaching the bosonic limit
where the superfluidity is suppressed.  As a result an increasingly more
attractive interaction is required to arrive at the shaded region.

A key observation of our theory is that the character of the $T_c$
curves changes at the point where $\mu$ changes sign.  Another key
observation is that, at large filling factors, the transition
temperature \textit{vanishes} near $\mu =0$, when the system 
approaches the bosonic regime. 
This vanishing is associated with a localization of pairs, that is, a
divergence of the pair effective mass.  Physically this is not
unexpected since we have seen the effective BEC Hamiltonian contains an
inter-site repulsion of the same magnitude as the hopping term. It is
this repulsion which inhibits pair hopping at large filling factors.
Only in the low density limit can this inter-site repulsion be
neglected, thereby leading to the conventional behavior, $T_c \propto
t^2/|U|$. At larger filling, we find that this simple BCS-Leggett ground
state will not support superfluidity in the $|U|\gg t$ limit.

We end this section by noting that while these calculations have been
based on a theory which is compatible with the BCS-Leggett ground state,
we believe our general conclusions will also apply if one were to
address BCS-BEC crossover in a Hubbard model using alternative crossover
schemes, for example, based on the NSR approach 
\cite{NSR}.  Below $T_c$, it has been shown \cite{PS05} that the NSR
scheme and related extensions treat the pairs in a weakly interacting
fashion.  This picture of weakly interacting ``bosons'' (at the level of
Bogoliubov or Popov theory) cannot be appropriate for the Hubbard model
near half filling and in the limit of strong attraction.  Consequently,
a proper self-consistent calculation of the transition temperature based
on this starting point should show signs of the inadequacy of the NSR
wave function ansatz, possibly through a vanishing of $T_c$ such as we
have found here.

\section{Numerical results for $d$-wave pairing}
\label{sec:dwave}

In this section we study the transition temperatures for the case of
$d$-wave pairing.  We can anticipate that as the system approaches
stronger coupling, pair hopping will be greatly suppressed, just as for
the $s$-wave case, and superfluidity will shut down. Indeed, because the
$d$-wave pairs are extended over two lattice sites, the effects we saw
for the $s$-wave case should be even more dramatic.  Moreover, because
there is no standard $d$-wave analogue of Eq.~(\ref{eq:Heff}), it is not
clear a priori whether the $\mu =0 $ point should reflect a qualitative
change in the physics.

\begin{figure} 
\centerline{\includegraphics[clip,width=3.4in]{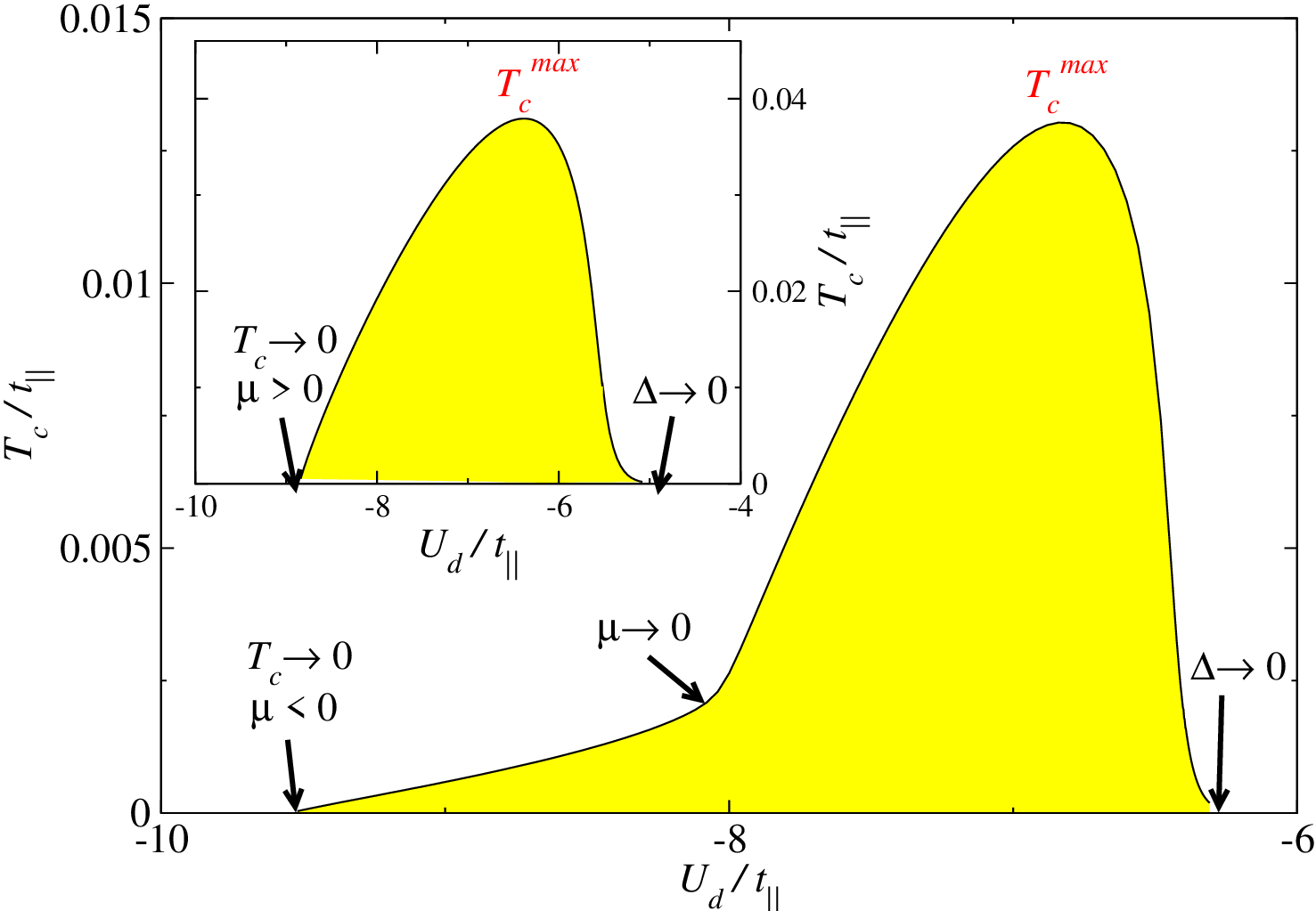}} 
\caption{(Color online) Transition temperature $T_c$ as a function of
  $U_d/t_\parallel$ for $d$-wave pairing on a quasi-2D lattice for
  $n=0.05$. At this low filling superfluidity can be sustained for some
  range of attractive interactions, even after $\mu$ becomes negative at
  moderately strong coupling.  For strong enough attraction we find a
  strictly zero $T_c$.  Inset: $T_c$ curve for $n=0.1$, for which $T_c$
  is nonzero only in the fermionic regime. }
\label{fig:dTc}
\end{figure}

Figure~\ref{fig:dTc} presents results for the $d$-wave transition
temperature in a quasi-2D lattice for $n=0.05$ and $n=0.1$ (inset). A
new feature emerges which is not present in the $s$-wave case. In the
ground state, there is a threshold in $|U_d|/t$ above which weak
coupling superfluidity is stable and below which it will not survive
\cite{Hertog}. Just as for the $s$-wave case, at all fillings there is a
maximum in the $T_c$ curves 
which we refer to as $T_c^{max}$, inside the fermionic regime. 
Importantly, only at extremely low filling ($n<0.1$) is $T_c$
finite when $\mu$ becomes negative. Otherwise $T_c$ vanishes for
sufficiently strong attraction, but still within the fermionic regime,
$\mu>0$. 
In fact, even for those extremely low values of $n$, where one can pass
into the bosonic regime, we find that $T_c $ reaches zero for
sufficiently large $|U_d|/t$.  This shut-down of superfluidity is
associated with a divergence in the pair mass, that is, with
localization of the pairs, just as for the $s$-wave case.

\begin{figure} 
\centerline{\includegraphics[clip,width=3.4in]{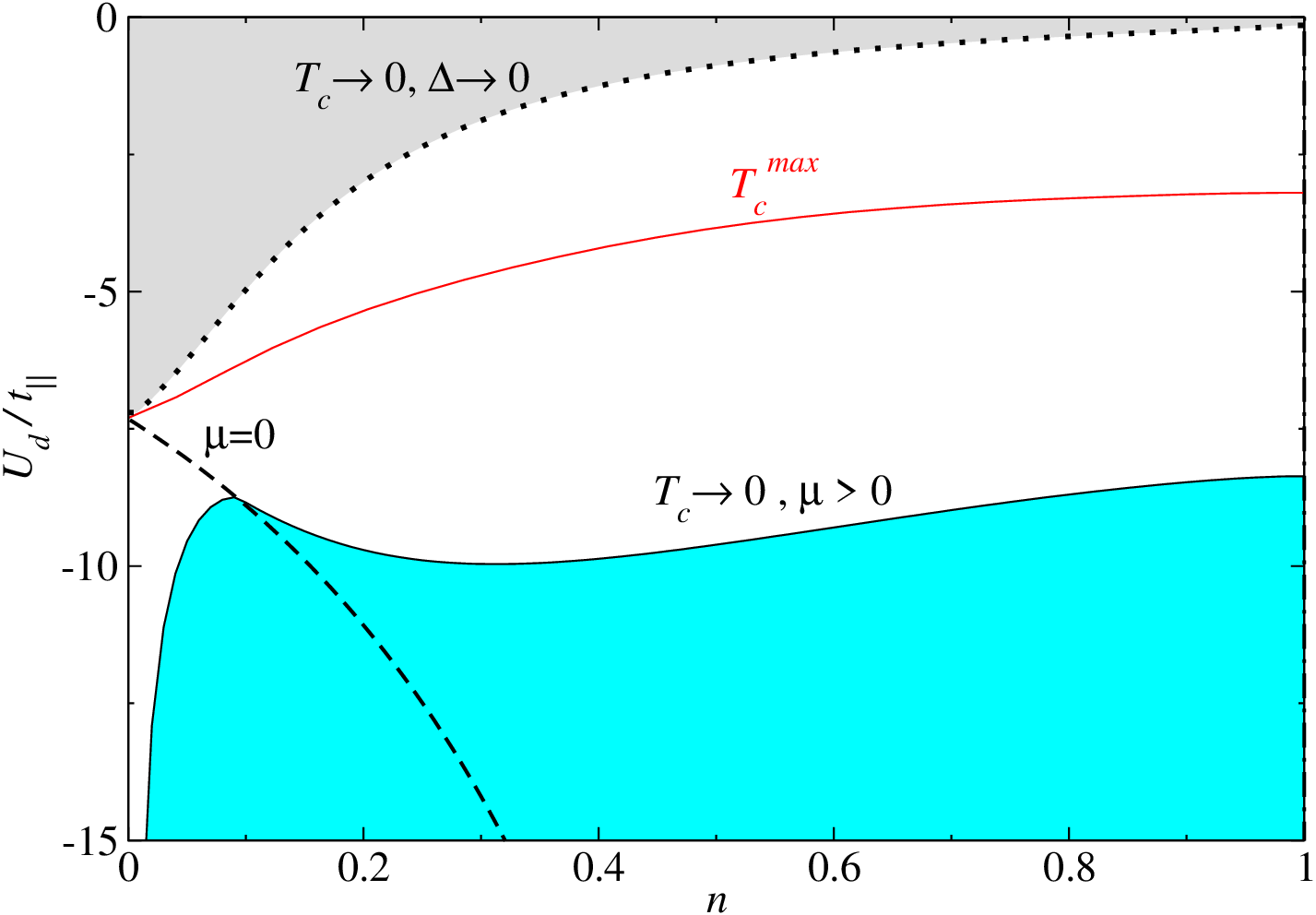}} 
\caption{(Color online) $T\rightarrow 0$ phase diagram for fermions on a
  quasi-2D lattice with $d$-wave pairing symmetry. The lower (cyan)
  shaded regimes shows where superfluidity does not survive as
  $T\rightarrow 0$. The black dotted line indicates the threshold beyond
  which superfluidity becomes observable (i.e., in the unshaded regions)
  and the upper (gray) shaded regime corresponds to a normal phase.  The
  red solid line shows the trace of $T_c^{max}$ shown in
  Fig.~\ref{fig:dTc}.  The black dash line indicates where $\mu=0$.}
\label{fig:dwaveT0}
\end{figure}

Shown in Fig.~\ref{fig:dwaveT0} is the phase diagram in the $U_d-n$
plane for the $d$-wave case. In the light (gray) shaded region, there is
no superfluid ground state associated with the BCS-Leggett $T=0$
equations \cite{Leggett}. This is on the weak coupling side of the phase
diagram.  In the dark (cyan) shaded region, pairing fluctuations destroy
superfluidity. This corresponds to the strong coupling side of the phase
diagram, and arises because of inhibition of hopping.  

It should be clear that in the $d$-wave case the regime where pairing
fluctuations destroy superfluidity is much larger than its counterpart
in the $s$-wave case. This reflects the fact that $d$-wave pairing
involves nearest-neighbor sites.  The finite size of the pairs (which
cannot be less than the lattice constant) distinguishes them from the
point-like composite bosons (of the $s$-wave case) and allows them to
break and recombine with neighboring sites in complicated ways. In the
very dilute limit we understand the behavior of Fig.~\ref{fig:dwaveT0}
as follows.  \textit{If} a two-body-like bound pair state exists (i.e.,
when $\mu < 0$), the binding energy will be given by $E_b \approx - 2
\mu \sim \hbar^2/2m\xi^2$, where $\xi$ is the pair size. Note that in a
many-body system, the dimensionless quantity $E_b/E_F$ must be large for
a system to be in the BEC limit.  The fact that the $d$-wave pair size
$\xi$ cannot be smaller than the lattice constant sets an upper bound
for the binding energy.

Only in the dilute limit where the mean inter-particle distance becomes
substantially larger than the $d$-wave pair size can one reach
sufficiently large values of $E_b/E_F$ to achieve a $d$-wave bosonic
superfluid, principally because of the small value of $n$ or $E_F$. In
this low density regime, if we continue to increase the attraction (at
very small fixed $n$) we again rapidly destabilize the superfluid
phase. This is associated with the fact that increased attraction
requires a larger and larger $E_b$ (or equivalently smaller $\xi$),
which eventually hits the limit set by the lattice constant.

Since the inter-particle distance $1/k_F$ in quasi-2D scales as
$1/\sqrt{E_F} \sim n^{-1/2}$, with increasing $n$ (at fixed attractive
interaction) the relative binding energy $E_b/E_F$ decreases very
rapidly. As a result, $1/k_F$ soon becomes comparable to the pair size
$\xi$, so that many-body effects strongly suppress the mobility of the
pairs. This eventually destroys the 
superfluid state.

Finally, as the density approaches half filling, additional effects arise
from particle-hole symmetry.  These tend to pin $\mu$ to
its noninteracting value ($\approx 4t_\parallel$). This
explains why the chemical potential is high when $T_c$ vanishes at
densities close to half filling. In summary, despite the fact that the
mean-field ground state equations have a superfluid solution, we find
that with fluctuation effects included, this $d$-wave superfluidity
(associated with the BCS-Leggett ground state) has difficulty becoming
established once the attractive interaction becomes moderately strong.

\section{Local pair state (LPS) wave function and BCS-LPS
  transition}

\subsection{The breakdown of the BCS-Leggett state in the $|U|\gg t$
  limit}

We have emphasized that the shaded region in Fig.~\ref{fig:L1T0} is
associated with the breakdown of the BCS-Leggett form of superfluidity.
It is natural to ask what is the nature of the phase inside the shaded
region.  We have in the past \cite{Chienlattice1} characterized this
state as ''insulating" on the basis of the general ``rule'' that a
bosonic system which is non-superfluid is generally localized. Moreover,
the onset of the shaded region that we find marks the onset of an
infinite pair mass which is consistent with localized bosons.  It has
been argued \cite{Oshikawa} that an insulating phase for both fermions
and bosons, if it corresponds to a finite excitation gap, is only
possible when the filling is commensurate. One way to get around these
arguments (which allow insulating phases away from commensurability) is
to introduce phase separation which we do not contemplate here.
It may also be that localized pair states with a
``localization
gap" arising from many body effects, (not disorder), cannot be ruled out.
Finally, one may also consider
other forms of superfluid phases which may be more stable than the
BCS-Leggett form, although they need not evolve continuously from the
BCS state.  Indeed, based on a study of the true Bose case
\cite{Scalettar} 
there is a suggestion that the ground state might well be a superfluid
whenever the nearest-neighbor repulsive term is smaller than twice the
hopping coefficient.

We observe that when the on-site attraction is strong, fermions are
expected to form local pairs and one might anticipate that a better
ground state for the superfluid phase is one where a given site either
has precisely zero ($n_i=0$) or precisely $2$ fermions.  Singly occupied
sites are unfavorable, although they do serve as opportunities for
virtual hopping.  It is convenient in what follows to count the number
of bosons which correspond to 1/2 of the fermion number. Due to the
particle-hole symmetry about $n_b=1/2$, we restrict consideration to
$0<n_b<(1/2)$ where $n_b=\sum_{i}\langle n_{b_{i}}\rangle /N$ is the
filling factor of pairs.

Following earlier work \cite{MicnasRMP}, we contemplate a new ground
state
\begin{equation}
|\mbox{LPS}\rangle=\prod_{i}(\sqrt{1-n_b}+\sqrt{n_b}b^{\dagger}_{i})|0\rangle.
\end{equation} 
An analogous wave function was discussed for bosonic systems
\cite{Rokhsar}. 
We stress that that the correct pair commutation relation
$[b_i,b^{\dagger}_i]=1-2n_{b_i}$ should be used. The ground state energy
for any $0<n_b\le(1/2)$ is $\langle
\mbox{LPS}|H_{eff}|\mbox{LPS}\rangle=Jzn_b(2n_b-1)N$. Here $z$ is the
number of nearest neighbors. Following \cite{Rokhsar} the
zero-momentum-pair 
 fraction is $\langle
LPS|b^{\dagger}_{i}|LPS\rangle^2/n_b=1-n_b$.  Thus, when $n_b\rightarrow
0$ nearly all pairs are in the zero-momentum condensate.

In this LPS state, near half filling, roughly half of the 
pairs at $T=0$ have finite momentum.  Note that this many body wave
function is associated with a macroscopic occupation of the lowest
effective single particle energy level, but that the effective single
particle levels need not be eigenstates of the non-interacting system.
Thus the ground state in question (for the strongly correlated case) is
not associated with ($\mathbf{k},-\mathbf{k}$) pairing. To see this in
more detail, we rewrite the LPS state in momentum space
\begin{eqnarray}
|LPS\rangle&=&\prod_{i}(\sqrt{1-n_b}+\sqrt{n_b}c^{\dagger}_{i\uparrow}c^{\dagger}_{i\downarrow})|0\rangle \\
&=&\prod_{i}(\sqrt{1-n_b}+\sqrt{n_b}\sum_{\mathbf{p},\mathbf{q}}c^{\dagger}_{\mathbf{p}\uparrow}c^{\dagger}_{\mathbf{q}\downarrow}e^{-i(\mathbf{p}+\mathbf{q})\cdot 
\mathbf{R}_{i}})|0\rangle. \nonumber 
\end{eqnarray} Here 
$\mathbf{R}_{i}$ denotes the position vector of the $i$-th site. From 
this expression one sees that when $n_b$ is finite, both finite momentum 
Cooper pairs 
$c^{\dagger}_{\mathbf{k}\uparrow}c^{\dagger}_{-\mathbf{k}
  +\mathbf{q}\downarrow}$ 
and higher order terms  such as 
$c^{\dagger}_{\mathbf{k}_1} c^{\dagger}_{\mathbf{k}_2}
c^{\dagger}_{\mathbf{k}_3} c^{\dagger}_{\mathbf{k}_4}$ 
are important. 
The presence of these finite momentum condensed pairs may relate to the
general issue of condensate fragmentation \cite{Fragmentation} which
occurs in the presence of degenerate ground states.  The degeneracy of
the strict atomic limit $|U|/t\rightarrow\infty$ is partially lifted
when weak tunneling is included, but it appears that a more natural
description of the superfluid phase should be one which abandons the
$(\mathbf{k}$, $-\mathbf{k})$ pairing of the BCS-Leggett phase.

We note that at low filling, one might expect that since the
nearest-neighbor repulsion is negligible this LPS ground state may not
be very different from the usual BCS-Leggett state. The
equivalence is straightforward to establish. For $n_b\ll 1$,
\begin{eqnarray}
|LPS\rangle&=&\prod_{i}(\sqrt{1-n_b}+\sqrt{n_b}b^{\dagger}_{i})|0\rangle
\nonumber \\
&\approx&(1+\sqrt{n_b}\sum_{\mathbf{k}}c^{\dagger}_{\mathbf{k}\uparrow}c^{\dagger}_{-\mathbf{k}\downarrow})|0\rangle
\nonumber \\
&\approx&\prod_{\mathbf{k}}(\sqrt{1-n_b}+\sqrt{n_b}c^{\dagger}_{\mathbf{k}\uparrow}c^{\dagger}_{-\mathbf{k}\downarrow})|0\rangle.
\end{eqnarray}
Here
$\sum_{i}b^{\dagger}_{i}=\sum_{i}c^{\dagger}_{i\uparrow}c^{\dagger}_{i\downarrow}=\sum_{\mathbf{k}}c^{\dagger}_{\mathbf{k}\uparrow}c^{\dagger}_{-\mathbf{k}\downarrow}$.
When $n_b\rightarrow 0$, the BCS number equation
$n=2n_b=2\sum_{\mathbf{k}}v^{2}_{\mathbf{k}}$ and the normalization
condition $u^{2}_{\mathbf{k}}+v^{2}_{\mathbf{k}}=1$ lead to
$u_{\mathbf{k}}\approx \sqrt{1-n_b}$ and
$v_{\mathbf{k}}\approx\sqrt{n_b}$.  As expected, the states
$|LPS\rangle$ and $|BCS\rangle$ are equivalent in the $|U|\gg t$
provided we consider the $n\rightarrow 0$ limit. This is consistent with
our earlier numerical results which show that there is no breakdown of
superfluidity at any strong attraction in the BCS-Leggett phase provided
the filling is low.  \textit{For small $n$}, the BCS wave function captures the
main features of the AHM in the two limits $|U|/t\ll 1$ and $|U|/t\gg 1$
and it should be appropriate for describing the crossover behavior of
the AHM for any $|U|/t$.
 This is no longer the case near half filling in the strongly attractive regime.
The BCS-Leggett wavefunction does not capture the physics associated with
strong inter-boson interactions, thereby leading us to contemplate
an alternate superfluid phase, such as that associated with the LPS wavefunction.
One might speculate that since the BCS-Leggett ground state and the LPS ground
state describe different types of superfluids for very different
limits, 
a quantum phase transition may occur when $|U|/t$ is tuned between these
two limits.  We speculate that the onset of this quantum phase
transition can be loosely associated with the boundary curve for the
shaded region in Fig.~\ref{fig:L1T0}.

\begin{figure}
\centerline{\includegraphics[clip,width=2.8in]{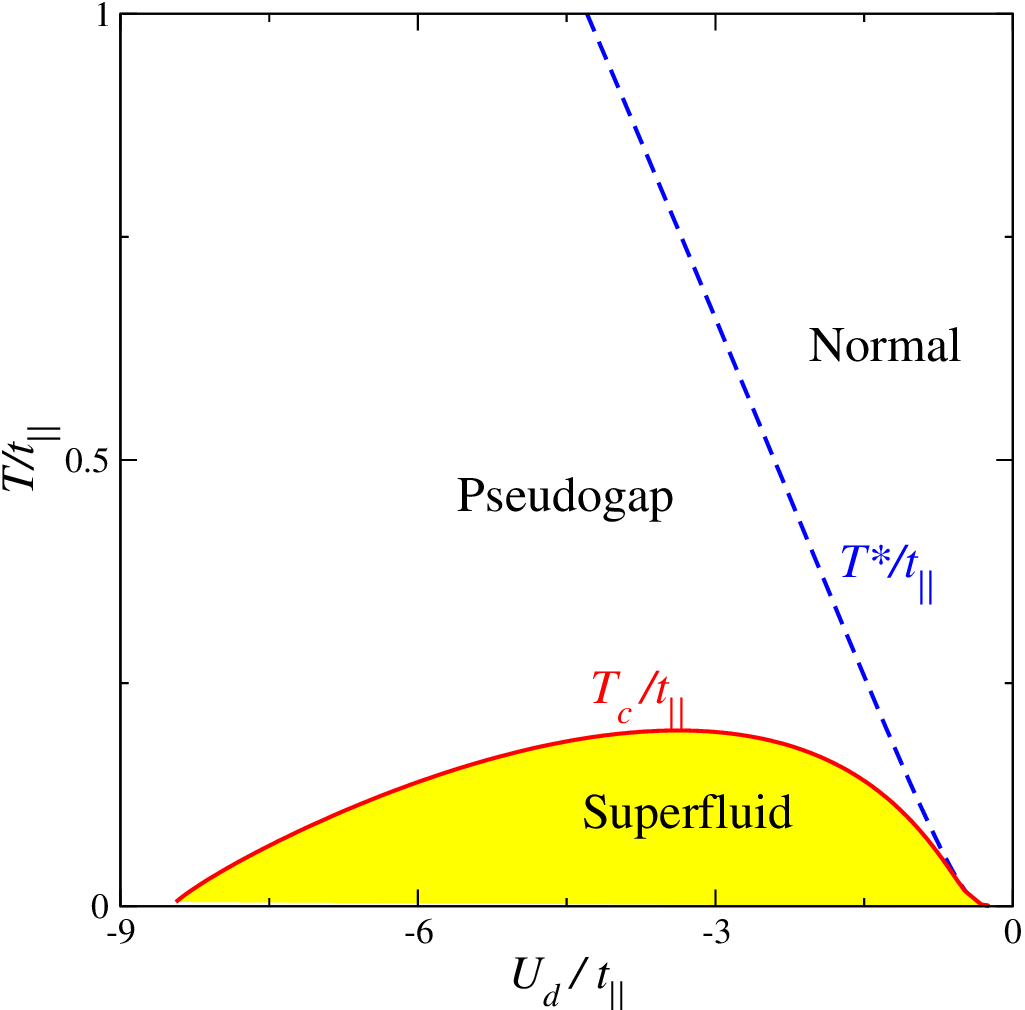}}
\caption{Phase diagram for $d$-wave pairing in a quasi-2D square lattice
  at $n=0.9$ (near half filling). Pairs emerge below $T^*$ but
  superfluidity only exists in the shaded regime. While the horizontal
  axis represents the strength of the attractive interaction, one may
  make contact with the cuprate phase diagram as shown in Fig.~7 of
  Ref.~\cite{LeeReview} by noting that the more underdoped the system
  the larger is $T^*$.  Thus, one may view the hole doping concentration
  $x=1-n$ as a means of parameterizing the pairing attraction.  In this
  way, there are some important similarities with the cuprates.}
\label{fig:dwaveTcTS}
\end{figure}

\section{Experimental Implications}

\subsection{Implications for High Temperature Superconductors}

While there is an extensive literature
\cite{Scalapino86,Pines1997,LeeReview} which argues that high $T_c$
superconductivity can be ``explained'' by the repulsive Hubbard model,
evidence that supports this point of view mainly comes from quantum
Monte Carlo simulation or numerical study using a density matrix
renormalization group method on a very small (e.g., 8x16) two
dimensional lattice.  Here we want to address what we can learn about
the cuprates from the AHM.  The phase diagram associated with the
attractive Hubbard case, importantly, extended to $d$-wave pairing, is
shown in Fig.~\ref{fig:dwaveTcTS} for the case of $n = 0.9$, which is
appropriate for the cuprates and corresponds to hole doping
concentration $x= 1-n = 0.1$.  The behavior of the transition
temperature was discussed in Sec.~\ref{sec:dwave}.  Here we add an
additional plot of $T^*$ which corresponds to the pairing onset
temperature. This is estimated by solving the standard mean field
equations at $\Delta = 0$.

We stress at the outset that the BCS-BEC crossover scenario
neither requires nor contains detailed microscopic information about the
pairing mechanism.  (More quantitative fits of this phase diagram to
that of the cuprates have been presented in the literature
\cite{ChenStripes}). Nevertheless, this phase diagram can be generically
compared to that of the cuprates (see, e.g., Fig.~7 of
Ref.~\cite{LeeReview}) without any detailed fits when we exploit the
fact that (in the cuprate data) as $x$ decreases, $T^*$ increases.  The
more underdoped the system the larger is $T^*$.  Thus, to make progress
we may view the hole doping concentration $x$ as a means of tuning the
size of the pairing attraction.  When $T^*$ is matched to the
experimental pseudogap onset temperature, the $d$-wave transition
temperature displays a maximum as in the inset of Fig.~\ref{fig:dTc},
just as seen experimentally at the optimal doping. Importantly, one sees
that when the attraction gets sufficiently strong superfluidity is shut
down.  Finally, we note that at the lower values for $T^*$ (that is, in
the overdoped regime), the superfluid phase in the cuprates is well
described by $d$-wave BCS theory and the pseudogap is negligible.  Thus
the end-point for superconductivity at the lower critical doping in the
cuprates may well be related to the break-down of BCS-BEC crossover
which we have been discussing in this paper. 
It should be stressed, however, that this disappearance of the $d$-wave
superfluid phase occurs well before one reaches the BEC limit. Indeed
Fig.~\ref{fig:dwaveT0} shows clearly that the fermionic chemical
potential is clearly positive (and large) when $T_c$ vanishes at high
fermion density. This is an important issue vis a vis the cuprates,
since it is clear that these systems have a large positive fermionic
chemical potential $\mu$ and are thus far from the bosonic regime.

A very important issue in high temperature superconductors, which has
not been resolved experimentally, is a determination of the nature of
that phase between a vanishing $T_c$ and a large $T^*$ at very low
doping concentrations where superconductivity disappears.  All that is
certain is that it corresponds to an insulating state
\cite{ourreview,LeeReview} in the sense that the resistivity appears to
increase with decreasing temperature. 
It should, in summary be clear that, although there may be differences
associated with long range Coulomb effects which may, for example lead
to phase separation \cite{Zachar1997}, studies of attractive
interactions on optical lattices have the potential for elucidating
important issues in the high $T_c$ cuprates.

\subsection{Experiments on Optical Lattices} 

In this section we discuss possible experiments to observe BEC-BCS
crossover on lattices in the specific context of the AHM.  It is not at
all straightforward to set up this fermionic one-band model and to avoid
introducing either multi-band effects or direct boson-boson hopping.  It
should be stressed that on lattices, BEC-BCS crossover need not rely on
a Feshbach resonance tuning of the attractive interaction alone.
Rather, to simulate the AHM, as shown in Refs.~\cite{Jaksch,Demler2,WernerPRL},
the on-site attraction $U$ and the hopping coefficient $t$ are
simultaneously tuned and are, thus, not independent.  They can generally
be controlled by varying the lattice potential depth $V_0$ as well as
the $s$-wave scattering length $a_s$ (of a Fermi gas in the absence of
lattice potentials). It is conventional to define the recoil energy as
$E_R\equiv h^2/(2m\lambda^2)$, where $\lambda$ is the wavelength of the
laser used to generate the lattice potential and the lattice spacing is
$d=\lambda/2$. Importantly, when $V_0\gg E_R$ and $d\gg |a_s|$, it can be
shown that the one-band Hubbard model emerges with
$t=(2/\sqrt{\pi})E_R\xi^3e^{-2\xi^2}$ and $U=\pi E_R(a_s/d)\xi^3$, where
$\xi=(V_0/E_R)^{1/4}$.

In the regime $a_s\ge d$ both multi-band effects and additional terms in
the Hamiltonian such as density-assisted hopping terms become important
\cite{WernerPRL}. Here the Hamiltonian can no longer be associated with
the AHM.  BCS-BEC crossover can readily occur in this limit without
the interruption we have found in our Hubbard calculations.  Indeed,
when a Fermi gas is in the strongly attractive regime (the BEC side of
a Feshbach resonance) in the absence of any lattice potential,
fermions form pairs with true binding energies
($\sim\hbar^2/2ma_s^2$), and pairs repel each other with an effective
scattering length $0.6a_s$ \cite{Petrov}. These two-body effects are
robust when an optical lattice is then ``switched on''. In this way
a
BEC limit can be readily established. More specifically,
one starts with a Fermi gas in the deep BEC regime
(satisfying $a_s\ll d$ when the lattice potential is turned off), and
then increases the lattice depth.  Importantly, the system can be
described by a Bose Hubbard model with infinite on-site repulsion. In
this case the two-body binding energy will dominate the on-site
attractive potential of fermions. There is no effective
nearest-neighbor repulsion (found in the BEC of fermion pairs) because
the true binding energy of pairs prevents the presence of unpaired
fermions and therefore eliminates these virtual processes.
It should be noted that the first generation experiments from MIT on
optical lattices \cite{MITlattice} were performed making use of a
Feshbach resonance so that before the optical lattice was established,
the system was near unitarity. Again, these experiments should not be
considered as simulating the one-band AHM.  Rather they pertain
to a model Hamiltonian which 
is clearly different from
the fermion Hubbard model. We have argued in this paper, as claimed
elsewhere
\cite{MicnasRMP}, that this attractive fermion Hubbard model may have
relevance to high temperature superconductors.

From the present perspective, based on the AHM, there are a variety of
important experiments yet to be done. Among these is to see if one can
find the predicted breakdown of BCS-BEC crossover which may be
associated with the transition to a new type of superfluid (The
BCS-Leggett state to the LPS state) or to some other type of
non-superfluid order.  To do this we propose that the Fermi gas is first
prepared in the weakly attractive regime (on the BCS side of resonance)
satisfying $|a_s|\ll d$ and then the lattice potential is gradually
increased to suppress hopping. As shown in Fig.~4 of
Ref.~\cite{WernerPRL}, this is equivalent to increasing $|U|/t$ in the
AHM when $a_s<0$. As $|U|/t$ increases, a maximum in $T_c$ should first
be observed. Near half filling when $|U|/t$ is sufficiently large (as
predicted by Fig.~\ref{fig:L1T0}) a new type of ordering may appear or
alternatively (if the associated transition temperatures are
sufficiently low) the system will be driven normal.  It is quite likely,
although it may be hard to verify, that the loss of superfluid order is
to be associated with a quantum phase transition near half filling and
when $|U|>>t$.

\section{Conclusions}

In this paper we have addressed a generalization of BCS-BEC crossover in
atomic Fermi systems to include the presence of an optical lattice. Our
specific interest is to address the AHM, and in the longer term, its
$d$-wave generalization which have been argued to be relevant to high
$T_c$ superconductors.  While there have been a number of numerical
studies \cite{Beck02,KellerTmatrix} of the attractive Hubbard
Hamiltonian, along with approaches based on dynamical mean field theory
\cite{KellerDMFT,WernerPRL}, ours is a systematic study over all filling
fractions and over the entire range of attractive interactions. 

We have
investigated the stability of the simplest type of superfluid phases
which represent a natural generalization of BCS-BEC crossover which has
been widely studied in the gas phases.  We refer to the ground state
wave function as the ``BCS-Leggett'' state and find that, even though
the strict $T = 0$ mean-field equations suggest that superfluidity can
occur everywhere, fluctuation effects lead to a vanishing $T_c$ over an
extended range near half filling and for moderately strong attraction.
Thus, there is an interruption of BCS-BEC crossover which we explain
from an analytic viewpoint. This interruption occurs very close to the
regime where the fermionic chemical potential $\mu =0$. It is a result
of the system passing over into the regime where the standard $t^2/U$-
expanded bosonic Hamiltonian (as an approximation to the AHM) becomes
valid. In contrast with the weakly interacting bosons of the BCS-Leggett
state (or of the NSR theory), here the bosonic degrees of freedom
experience a strong inter-site repulsion. Importantly 
superfluidity in these specific (weak coupling) forms cannot be
supported.  

We posit an alternative superfluid phase which may be more
appropriate at very strong attraction.
These studies have been extended to the $d$-wave case as well.  We
stress that the resulting phase diagram has many features in common with
that of the cuprates.  In particular, at sufficiently strong attraction
(as represented by a large value for the pairing onset temperature,
called $T^*$), superfluidity is interrupted.  This interruption of
BCS-BEC crossover occurs while the system still has a rather large
positive fermionic chemical potential, and hence is far from the BEC.

A central goal of this body of work is to make a case for future optical
lattice experiments to simulate the AHM, first for the $s$-wave and
ultimately for the $d$-wave case. It is our contention that in this way
we have as much to learn about the cuprate superconductors, as from
studies of the repulsive Hubbard model which has been conjectured to
give rise to $d$-wave attraction.

This work is supported by NSF PHY-0555325 and NSF-MRSEC Grant
No.~DMR-0213745. We thank Cheng Chin for stimulating discussions and A.
Paramekanti for helpful insight.

\vspace*{-1ex} \bibliographystyle{apsrev}


\begin{thebibliography}{50}
\expandafter\ifx\csname natexlab\endcsname\relax\def\natexlab#1{#1}\fi
\expandafter\ifx\csname bibnamefont\endcsname\relax
  \def\bibnamefont#1{#1}\fi
\expandafter\ifx\csname bibfnamefont\endcsname\relax
  \def\bibfnamefont#1{#1}\fi
\expandafter\ifx\csname citenamefont\endcsname\relax
  \def\citenamefont#1{#1}\fi
\expandafter\ifx\csname url\endcsname\relax
  \def\url#1{\texttt{#1}}\fi
\expandafter\ifx\csname urlprefix\endcsname\relax\def\urlprefix{URL }\fi
\providecommand{\bibinfo}[2]{#2}
\providecommand{\eprint}[2][]{\url{#2}}

\bibitem[{\citenamefont{Stoferle et~al.}(2006)\citenamefont{Stoferle, Moritz,
  Gunter, Kohl, and Esslinger}}]{Stoferle}
\bibinfo{author}{\bibfnamefont{T.}~\bibnamefont{Stoferle}},
  \bibinfo{author}{\bibfnamefont{H.}~\bibnamefont{Moritz}},
  \bibinfo{author}{\bibfnamefont{K.}~\bibnamefont{Gunter}},
  \bibinfo{author}{\bibfnamefont{M.}~\bibnamefont{Kohl}}, \bibnamefont{and}
  \bibinfo{author}{\bibfnamefont{T.}~\bibnamefont{Esslinger}},
  \bibinfo{journal}{Phys. Rev. Lett.} \textbf{\bibinfo{volume}{96}},
  \bibinfo{pages}{030401} (\bibinfo{year}{2006}).

\bibitem[{\citenamefont{Chin et~al.}(2006)\citenamefont{Chin, Miller, Liu,
  Stan, Setiawan, Sanner, Xu, and Ketterle}}]{MITlattice}
\bibinfo{author}{\bibfnamefont{J.~K.} \bibnamefont{Chin}},
  \bibinfo{author}{\bibfnamefont{D.~E.} \bibnamefont{Miller}},
  \bibinfo{author}{\bibfnamefont{Y.}~\bibnamefont{Liu}},
  \bibinfo{author}{\bibfnamefont{C.}~\bibnamefont{Stan}},
  \bibinfo{author}{\bibfnamefont{W.}~\bibnamefont{Setiawan}},
  \bibinfo{author}{\bibfnamefont{C.}~\bibnamefont{Sanner}},
  \bibinfo{author}{\bibfnamefont{K.}~\bibnamefont{Xu}}, \bibnamefont{and}
  \bibinfo{author}{\bibfnamefont{W.}~\bibnamefont{Ketterle}},
  \bibinfo{journal}{Nature} \textbf{\bibinfo{volume}{443}},
  \bibinfo{pages}{961} (\bibinfo{year}{2006}).

\bibitem[{\citenamefont{Hofstetter et~al.}(2002)\citenamefont{Hofstetter,
  Cirac, Zoller, Demler, and Lukin}}]{Demler2}
\bibinfo{author}{\bibfnamefont{W.}~\bibnamefont{Hofstetter}},
  \bibinfo{author}{\bibfnamefont{J.~I.} \bibnamefont{Cirac}},
  \bibinfo{author}{\bibfnamefont{P.}~\bibnamefont{Zoller}},
  \bibinfo{author}{\bibfnamefont{E.}~\bibnamefont{Demler}}, \bibnamefont{and}
  \bibinfo{author}{\bibfnamefont{M.~D.} \bibnamefont{Lukin}},
  \bibinfo{journal}{Phys. Rev. Lett.} \textbf{\bibinfo{volume}{89}},
  \bibinfo{pages}{220407} (\bibinfo{year}{2002}).

\bibitem[{\citenamefont{Georges}(2007)}]{GeorgesReview}
\bibinfo{author}{\bibfnamefont{A.}~\bibnamefont{Georges}}, in
  \emph{\bibinfo{booktitle}{Ultra-cold Fermi Gases}}, edited by
  \bibinfo{editor}{\bibfnamefont{M.}~\bibnamefont{Inguscio}},
  \bibinfo{editor}{\bibfnamefont{W.}~\bibnamefont{Ketterle}}, \bibnamefont{and}
  \bibinfo{editor}{\bibfnamefont{C.}~\bibnamefont{Salomon}}
  (\bibinfo{organization}{Italian physical society}, \bibinfo{year}{2007}), p.
  \bibinfo{pages}{477}.

\bibitem[{\citenamefont{Leggett}(1980)}]{Leggett}
\bibinfo{author}{\bibfnamefont{A.~J.} \bibnamefont{Leggett}}, in
  \emph{\bibinfo{booktitle}{Modern Trends in the Theory of Condensed Matter}}
  (\bibinfo{publisher}{Springer-Verlag}, \bibinfo{address}{Berlin},
  \bibinfo{year}{1980}), pp. \bibinfo{pages}{13--27}.

\bibitem[{\citenamefont{Nozi\`{e}res and Schmitt-Rink}(1985)}]{NSR}
\bibinfo{author}{\bibfnamefont{P.}~\bibnamefont{Nozi\`{e}res}}
  \bibnamefont{and}
  \bibinfo{author}{\bibfnamefont{S.}~\bibnamefont{Schmitt-Rink}},
  \bibinfo{journal}{J. Low Temp. Phys.} \textbf{\bibinfo{volume}{59}},
  \bibinfo{pages}{195} (\bibinfo{year}{1985}).

\bibitem[{\citenamefont{Randeria}(1995)}]{Randeriareview}
\bibinfo{author}{\bibfnamefont{M.}~\bibnamefont{Randeria}}, in
  \emph{\bibinfo{booktitle}{Bose Einstein Condensation}}, edited by
  \bibinfo{editor}{\bibfnamefont{A.}~\bibnamefont{Griffin}},
  \bibinfo{editor}{\bibfnamefont{D.}~\bibnamefont{Snoke}}, \bibnamefont{and}
  \bibinfo{editor}{\bibfnamefont{S.}~\bibnamefont{Stringari}}
  (\bibinfo{publisher}{Cambridge Univ. Press}, \bibinfo{address}{Cambridge},
  \bibinfo{year}{1995}), pp. \bibinfo{pages}{355--92}.

\bibitem[{\citenamefont{Lee et~al.}(2006)\citenamefont{Lee, Nagaosa, and
  Wen}}]{LeeReview}
\bibinfo{author}{\bibfnamefont{P.~A.} \bibnamefont{Lee}},
  \bibinfo{author}{\bibfnamefont{N.}~\bibnamefont{Nagaosa}}, \bibnamefont{and}
  \bibinfo{author}{\bibfnamefont{X.~G.} \bibnamefont{Wen}},
  \bibinfo{journal}{Rev. Mod. Phys.} \textbf{\bibinfo{volume}{78}},
  \bibinfo{pages}{17} (\bibinfo{year}{2006}).

\bibitem[{\citenamefont{Pines}(1997)}]{Pines1997}
\bibinfo{author}{\bibfnamefont{D.}~\bibnamefont{Pines}},
  \bibinfo{journal}{Physica C} \textbf{\bibinfo{volume}{282-287}},
  \bibinfo{pages}{273} (\bibinfo{year}{1997}).

\bibitem[{\citenamefont{Scalapino et~al.}(1986)\citenamefont{Scalapino, Loh,
  and Hirsch}}]{Scalapino86}
\bibinfo{author}{\bibfnamefont{D.~J.} \bibnamefont{Scalapino}},
  \bibinfo{author}{\bibfnamefont{E.}~\bibnamefont{Loh}}, \bibnamefont{and}
  \bibinfo{author}{\bibfnamefont{J.~E.} \bibnamefont{Hirsch}},
  \bibinfo{journal}{Phys. Rev. B} \textbf{\bibinfo{volume}{34}},
  \bibinfo{pages}{8190} (\bibinfo{year}{1986}).

\bibitem[{\citenamefont{Chen et~al.}(2005)\citenamefont{Chen, Stajic, Tan, and
  Levin}}]{ourreview}
\bibinfo{author}{\bibfnamefont{Q.~J.} \bibnamefont{Chen}},
  \bibinfo{author}{\bibfnamefont{J.}~\bibnamefont{Stajic}},
  \bibinfo{author}{\bibfnamefont{S.~N.} \bibnamefont{Tan}}, \bibnamefont{and}
  \bibinfo{author}{\bibfnamefont{K.}~\bibnamefont{Levin}},
  \bibinfo{journal}{Phys. Rep.} \textbf{\bibinfo{volume}{412}},
  \bibinfo{pages}{1} (\bibinfo{year}{2005}).

\bibitem[{\citenamefont{Giorgini et~al.}(2007)\citenamefont{Giorgini,
  Pitaevskii, and Stringari}}]{StringariReview}
\bibinfo{author}{\bibfnamefont{S.}~\bibnamefont{Giorgini}},
  \bibinfo{author}{\bibfnamefont{L.~P.} \bibnamefont{Pitaevskii}},
  \bibnamefont{and} \bibinfo{author}{\bibfnamefont{S.}~\bibnamefont{Stringari}}
  (\bibinfo{year}{2007}), \bibinfo{note}{eprint, arXiv:0706.3360}.

\bibitem[{\citenamefont{Leggett}(2006)}]{LeggettNature}
\bibinfo{author}{\bibfnamefont{A.~J.} \bibnamefont{Leggett}},
  \bibinfo{journal}{Nature Physics} \textbf{\bibinfo{volume}{2}},
  \bibinfo{pages}{134} (\bibinfo{year}{2006}).

\bibitem[{\citenamefont{Levin and Chen}(2007)}]{Varenna}
\bibinfo{author}{\bibfnamefont{K.}~\bibnamefont{Levin}} \bibnamefont{and}
  \bibinfo{author}{\bibfnamefont{Q.}~\bibnamefont{Chen}}, in
  \emph{\bibinfo{booktitle}{Ultra-cold Fermi Gases}}, edited by
  \bibinfo{editor}{\bibfnamefont{M.}~\bibnamefont{Inguscio}},
  \bibinfo{editor}{\bibfnamefont{W.}~\bibnamefont{Ketterle}}, \bibnamefont{and}
  \bibinfo{editor}{\bibfnamefont{C.}~\bibnamefont{Salomon}}
  (\bibinfo{organization}{Italian physical society}, \bibinfo{year}{2007}), p.
  \bibinfo{pages}{751}.

\bibitem[{\citenamefont{Liu and Levin}(1997)}]{Liu}
\bibinfo{author}{\bibfnamefont{D.~Z.} \bibnamefont{Liu}} \bibnamefont{and}
  \bibinfo{author}{\bibfnamefont{K.}~\bibnamefont{Levin}},
  \bibinfo{journal}{Physica C} \textbf{\bibinfo{volume}{275}},
  \bibinfo{pages}{81} (\bibinfo{year}{1997}).

\bibitem[{\citenamefont{Leggett}(1999)}]{Leggett2}
\bibinfo{author}{\bibfnamefont{A.~J.} \bibnamefont{Leggett}},
  \bibinfo{journal}{Phys. Rev. Lett.} \textbf{\bibinfo{volume}{83}},
  \bibinfo{pages}{392} (\bibinfo{year}{1999}).

\bibitem[{\citenamefont{Keller et~al.}(1999)\citenamefont{Keller, Metzner, and
  Schollwock}}]{KellerTmatrix}
\bibinfo{author}{\bibfnamefont{M.}~\bibnamefont{Keller}},
  \bibinfo{author}{\bibfnamefont{W.}~\bibnamefont{Metzner}}, \bibnamefont{and}
  \bibinfo{author}{\bibfnamefont{U.}~\bibnamefont{Schollwock}},
  \bibinfo{journal}{Phys. Rev. B} \textbf{\bibinfo{volume}{60}},
  \bibinfo{pages}{3499} (\bibinfo{year}{1999}).

\bibitem[{\citenamefont{Sewer et~al.}(2002)\citenamefont{Sewer, Zotos, and
  Beck}}]{Beck02}
\bibinfo{author}{\bibfnamefont{A.}~\bibnamefont{Sewer}},
  \bibinfo{author}{\bibfnamefont{X.}~\bibnamefont{Zotos}}, \bibnamefont{and}
  \bibinfo{author}{\bibfnamefont{H.}~\bibnamefont{Beck}},
  \bibinfo{journal}{Phys. Rev. B} \textbf{\bibinfo{volume}{66}},
  \bibinfo{pages}{140504(R)} (\bibinfo{year}{2002}).

\bibitem[{\citenamefont{Paiva et~al.}(2004)\citenamefont{Paiva, dos Santos,
  Scalettar, and Denteneer}}]{Scalettar2D}
\bibinfo{author}{\bibfnamefont{T.}~\bibnamefont{Paiva}},
  \bibinfo{author}{\bibfnamefont{R.~R.} \bibnamefont{dos Santos}},
  \bibinfo{author}{\bibfnamefont{R.~T.} \bibnamefont{Scalettar}},
  \bibnamefont{and} \bibinfo{author}{\bibfnamefont{P.~J.~H.}
  \bibnamefont{Denteneer}}, \bibinfo{journal}{Phys. Rev. B}
  \textbf{\bibinfo{volume}{69}}, \bibinfo{pages}{184501}
  (\bibinfo{year}{2004}).

\bibitem[{\citenamefont{Keller et~al.}(2001)\citenamefont{Keller, Metzner, and
  Schollwock}}]{KellerDMFT}
\bibinfo{author}{\bibfnamefont{M.}~\bibnamefont{Keller}},
  \bibinfo{author}{\bibfnamefont{W.}~\bibnamefont{Metzner}}, \bibnamefont{and}
  \bibinfo{author}{\bibfnamefont{U.}~\bibnamefont{Schollwock}},
  \bibinfo{journal}{Phys. Rev. Lett.} \textbf{\bibinfo{volume}{86}},
  \bibinfo{pages}{4612} (\bibinfo{year}{2001}).

\bibitem[{\citenamefont{Toschi et~al.}(2005{\natexlab{a}})\citenamefont{Toschi,
  Barone, Capone, and Castellani}}]{CaponeDMFT1}
\bibinfo{author}{\bibfnamefont{A.}~\bibnamefont{Toschi}},
  \bibinfo{author}{\bibfnamefont{P.}~\bibnamefont{Barone}},
  \bibinfo{author}{\bibfnamefont{M.}~\bibnamefont{Capone}}, \bibnamefont{and}
  \bibinfo{author}{\bibfnamefont{C.}~\bibnamefont{Castellani}},
  \bibinfo{journal}{New J. Phys.} \textbf{\bibinfo{volume}{7}},
  \bibinfo{pages}{7} (\bibinfo{year}{2005}{\natexlab{a}}).

\bibitem[{\citenamefont{Toschi et~al.}(2005{\natexlab{b}})\citenamefont{Toschi,
  Capone, and Castellani}}]{CaponeDMFT2}
\bibinfo{author}{\bibfnamefont{A.}~\bibnamefont{Toschi}},
  \bibinfo{author}{\bibfnamefont{M.}~\bibnamefont{Capone}}, \bibnamefont{and}
  \bibinfo{author}{\bibfnamefont{C.}~\bibnamefont{Castellani}},
  \bibinfo{journal}{Phys. Rev. B} \textbf{\bibinfo{volume}{72}},
  \bibinfo{pages}{235118} (\bibinfo{year}{2005}{\natexlab{b}}).

\bibitem[{\citenamefont{Werner et~al.}(2005)\citenamefont{Werner, Parcollet,
  Georges, and Hassan}}]{WernerPRL}
\bibinfo{author}{\bibfnamefont{F.}~\bibnamefont{Werner}},
  \bibinfo{author}{\bibfnamefont{O.}~\bibnamefont{Parcollet}},
  \bibinfo{author}{\bibfnamefont{A.}~\bibnamefont{Georges}}, \bibnamefont{and}
  \bibinfo{author}{\bibfnamefont{S.~R.} \bibnamefont{Hassan}},
  \bibinfo{journal}{Phys. Rev. Lett.} \textbf{\bibinfo{volume}{95}},
  \bibinfo{pages}{056401} (\bibinfo{year}{2005}).

\bibitem[{\citenamefont{Tan and Levin}(2006)}]{Shina2}
\bibinfo{author}{\bibfnamefont{S.~N.} \bibnamefont{Tan}} \bibnamefont{and}
  \bibinfo{author}{\bibfnamefont{K.}~\bibnamefont{Levin}},
  \bibinfo{journal}{\pra} \textbf{\bibinfo{volume}{74}},
  \bibinfo{pages}{043606} (\bibinfo{year}{2006}).

\bibitem[{\citenamefont{Chien et~al.}(2008)\citenamefont{Chien, He, Chen, and
  Levin}}]{Chienlattice1}
\bibinfo{author}{\bibfnamefont{C.-C.} \bibnamefont{Chien}},
  \bibinfo{author}{\bibfnamefont{Y.}~\bibnamefont{He}},
  \bibinfo{author}{\bibfnamefont{Q.}~\bibnamefont{Chen}}, \bibnamefont{and}
  \bibinfo{author}{\bibfnamefont{K.}~\bibnamefont{Levin}},
  \bibinfo{journal}{Phys. Rev. A} \textbf{\bibinfo{volume}{77}},
  \bibinfo{pages}{011601(R)} (\bibinfo{year}{2008}).

\bibitem[{\citenamefont{Koetsier et~al.}(2006)\citenamefont{Koetsier,
  Dickerscheid, and Stoof}}]{Stooflattice}
\bibinfo{author}{\bibfnamefont{A.}~\bibnamefont{Koetsier}},
  \bibinfo{author}{\bibfnamefont{D.~B.~M.} \bibnamefont{Dickerscheid}},
  \bibnamefont{and} \bibinfo{author}{\bibfnamefont{H.~T.~C.}
  \bibnamefont{Stoof}}, \bibinfo{journal}{Phys. Rev. A}
  \textbf{\bibinfo{volume}{74}}, \bibinfo{pages}{033621}
  (\bibinfo{year}{2006}).

\bibitem[{\citenamefont{Fedichev et~al.}(2004)\citenamefont{Fedichev, Bijlsma,
  and Zoller}}]{Fedichev}
\bibinfo{author}{\bibfnamefont{P.~O.} \bibnamefont{Fedichev}},
  \bibinfo{author}{\bibfnamefont{M.~J.} \bibnamefont{Bijlsma}},
  \bibnamefont{and} \bibinfo{author}{\bibfnamefont{P.}~\bibnamefont{Zoller}},
  \bibinfo{journal}{Phys. Rev. Lett.} \textbf{\bibinfo{volume}{92}},
  \bibinfo{pages}{080401} (\bibinfo{year}{2004}).

\bibitem[{\citenamefont{Micnas et~al.}(1990)\citenamefont{Micnas, Ranninger,
  and Robaszkiewicz}}]{MicnasRMP}
\bibinfo{author}{\bibfnamefont{R.}~\bibnamefont{Micnas}},
  \bibinfo{author}{\bibfnamefont{J.}~\bibnamefont{Ranninger}},
  \bibnamefont{and}
  \bibinfo{author}{\bibfnamefont{S.}~\bibnamefont{Robaszkiewicz}},
  \bibinfo{journal}{Rev. Mod. Phys.} \textbf{\bibinfo{volume}{62}},
  \bibinfo{pages}{113} (\bibinfo{year}{1990}).

\bibitem[{\citenamefont{Carr and Holland}(2005)}]{Carr05}
\bibinfo{author}{\bibfnamefont{L.~D.} \bibnamefont{Carr}} \bibnamefont{and}
  \bibinfo{author}{\bibfnamefont{M.~J.} \bibnamefont{Holland}},
  \bibinfo{journal}{Phys. Rev. A} \textbf{\bibinfo{volume}{72}},
  \bibinfo{pages}{031604(R)} (\bibinfo{year}{2005}).

\bibitem[{\citenamefont{Zhou}(2005)}]{Zhou05}
\bibinfo{author}{\bibfnamefont{F.}~\bibnamefont{Zhou}}, \bibinfo{journal}{Phys.
  Rev. B} \textbf{\bibinfo{volume}{72}}, \bibinfo{pages}{220501(R)}
  (\bibinfo{year}{2005}).

\bibitem[{\citenamefont{Zhai and Ho}(2007)}]{Holattice}
\bibinfo{author}{\bibfnamefont{H.}~\bibnamefont{Zhai}} \bibnamefont{and}
  \bibinfo{author}{\bibfnamefont{T.~L.} \bibnamefont{Ho}},
  \bibinfo{journal}{Phys. Rev. Lett.} \textbf{\bibinfo{volume}{99}},
  \bibinfo{pages}{100402} (\bibinfo{year}{2007}).

\bibitem[{\citenamefont{Moon et~al.}(2007)\citenamefont{Moon, Nikolic, and
  Sachdev}}]{Nikolic}
\bibinfo{author}{\bibfnamefont{E.~G.} \bibnamefont{Moon}},
  \bibinfo{author}{\bibfnamefont{P.}~\bibnamefont{Nikolic}}, \bibnamefont{and}
  \bibinfo{author}{\bibfnamefont{S.}~\bibnamefont{Sachdev}},
  \bibinfo{journal}{Phys. Rev. Lett.} \textbf{\bibinfo{volume}{99}},
  \bibinfo{pages}{230403} (\bibinfo{year}{2007}).

\bibitem[{\citenamefont{Burkov and Paramekanti}(2008)}]{Burkov}
\bibinfo{author}{\bibfnamefont{A.~A.} \bibnamefont{Burkov}} \bibnamefont{and}
  \bibinfo{author}{\bibfnamefont{A.}~\bibnamefont{Paramekanti}},
  \bibinfo{journal}{Phys. Rev. Lett.} \textbf{\bibinfo{volume}{100}},
  \bibinfo{pages}{255301} (\bibinfo{year}{2008}).

\bibitem[{\citenamefont{Lieb}(1989)}]{Lieb}
\bibinfo{author}{\bibfnamefont{E.~H.} \bibnamefont{Lieb}},
  \bibinfo{journal}{Phys. Rev. Lett.} \textbf{\bibinfo{volume}{62}},
  \bibinfo{pages}{1201} (\bibinfo{year}{1989}).

\bibitem[{\citenamefont{Chen et~al.}(1999)\citenamefont{Chen, Kosztin, Jank\'o,
  and Levin}}]{Chen1}
\bibinfo{author}{\bibfnamefont{Q.~J.} \bibnamefont{Chen}},
  \bibinfo{author}{\bibfnamefont{I.}~\bibnamefont{Kosztin}},
  \bibinfo{author}{\bibfnamefont{B.}~\bibnamefont{Jank\'o}}, \bibnamefont{and}
  \bibinfo{author}{\bibfnamefont{K.}~\bibnamefont{Levin}},
  \bibinfo{journal}{Phys. Rev. B} \textbf{\bibinfo{volume}{59}},
  \bibinfo{pages}{7083} (\bibinfo{year}{1999}).

\bibitem[{\citenamefont{Micnas}(2007)}]{Micnaslattice}
\bibinfo{author}{\bibfnamefont{R.}~\bibnamefont{Micnas}},
  \bibinfo{journal}{Phys. Rev. B} \textbf{\bibinfo{volume}{76}},
  \bibinfo{pages}{184507} (\bibinfo{year}{2007}).

\bibitem[{\citenamefont{Fisher et~al.}(1989)\citenamefont{Fisher, Weichman,
  Grinstein, and Fisher}}]{MPFisher}
\bibinfo{author}{\bibfnamefont{M.~P.~A.} \bibnamefont{Fisher}},
  \bibinfo{author}{\bibfnamefont{P.~B.} \bibnamefont{Weichman}},
  \bibinfo{author}{\bibfnamefont{G.}~\bibnamefont{Grinstein}},
  \bibnamefont{and} \bibinfo{author}{\bibfnamefont{D.~S.}
  \bibnamefont{Fisher}}, \bibinfo{journal}{Phys. Rev. B}
  \textbf{\bibinfo{volume}{40}}, \bibinfo{pages}{546} (\bibinfo{year}{1989}).

\bibitem[{\citenamefont{Rokhsar and Kotliar}(1991)}]{Rokhsar}
\bibinfo{author}{\bibfnamefont{D.~S.} \bibnamefont{Rokhsar}} \bibnamefont{and}
  \bibinfo{author}{\bibfnamefont{B.~G.} \bibnamefont{Kotliar}},
  \bibinfo{journal}{Phys. Rev. B} \textbf{\bibinfo{volume}{44}},
  \bibinfo{pages}{10328} (\bibinfo{year}{1991}).

\bibitem[{\citenamefont{Scalettar et~al.}(1995)\citenamefont{Scalettar,
  Batrouni, Kampf, and Zimanyi}}]{Scalettar}
\bibinfo{author}{\bibfnamefont{R.~T.} \bibnamefont{Scalettar}},
  \bibinfo{author}{\bibfnamefont{G.~G.} \bibnamefont{Batrouni}},
  \bibinfo{author}{\bibfnamefont{A.~P.} \bibnamefont{Kampf}}, \bibnamefont{and}
  \bibinfo{author}{\bibfnamefont{G.~T.} \bibnamefont{Zimanyi}},
  \bibinfo{journal}{Phys. Rev. B} \textbf{\bibinfo{volume}{51}},
  \bibinfo{pages}{8467} (\bibinfo{year}{1995}).

\bibitem[{\citenamefont{den Hertog}(1999)}]{Hertog}
\bibinfo{author}{\bibfnamefont{B.~C.} \bibnamefont{den Hertog}},
  \bibinfo{journal}{Phys. Rev. B} \textbf{\bibinfo{volume}{60}},
  \bibinfo{pages}{559} (\bibinfo{year}{1999}).

\bibitem[{\citenamefont{Robaszkiewicz et~al.}(1981)\citenamefont{Robaszkiewicz,
  Micnas, and Chao}}]{Rob}
\bibinfo{author}{\bibfnamefont{S.~R.} \bibnamefont{Robaszkiewicz}},
  \bibinfo{author}{\bibfnamefont{R.}~\bibnamefont{Micnas}}, \bibnamefont{and}
  \bibinfo{author}{\bibfnamefont{K.~A.} \bibnamefont{Chao}},
  \bibinfo{journal}{Phys. Rev. B} \textbf{\bibinfo{volume}{23}},
  \bibinfo{pages}{1447} (\bibinfo{year}{1981}).

\bibitem[{\citenamefont{Schrieffer}(1964)}]{S64}
\bibinfo{author}{\bibfnamefont{J.~R.} \bibnamefont{Schrieffer}},
  \emph{\bibinfo{title}{Theory of Superconductivity}}
  (\bibinfo{publisher}{Benjamin}, \bibinfo{address}{New York},
  \bibinfo{year}{1964}).

\bibitem[{\citenamefont{Bloch et~al.}(2007)\citenamefont{Bloch, Dalibard, and
  Zwerger}}]{BlochRMP}
\bibinfo{author}{\bibfnamefont{I.}~\bibnamefont{Bloch}},
  \bibinfo{author}{\bibfnamefont{J.}~\bibnamefont{Dalibard}}, \bibnamefont{and}
  \bibinfo{author}{\bibfnamefont{W.}~\bibnamefont{Zwerger}}
  (\bibinfo{year}{2007}), \bibinfo{note}{arXiv:0704.3011}.

\bibitem[{\citenamefont{Pieri and Strinati}(2005)}]{PS05}
\bibinfo{author}{\bibfnamefont{P.}~\bibnamefont{Pieri}} \bibnamefont{and}
  \bibinfo{author}{\bibfnamefont{G.~C.} \bibnamefont{Strinati}},
  \bibinfo{journal}{Phys. Rev. B.} \textbf{\bibinfo{volume}{71}},
  \bibinfo{pages}{094520} (\bibinfo{year}{2005}).

\bibitem[{\citenamefont{Oshikawa}(2000)}]{Oshikawa}
\bibinfo{author}{\bibfnamefont{M.}~\bibnamefont{Oshikawa}},
  \bibinfo{journal}{Phys. Rev. Lett.} \textbf{\bibinfo{volume}{84}},
  \bibinfo{pages}{1535} (\bibinfo{year}{2000}).

\bibitem[{\citenamefont{Mueller et~al.}(2006)\citenamefont{Mueller, Ho, Ueda,
  and Baym}}]{Fragmentation}
\bibinfo{author}{\bibfnamefont{E.~J.} \bibnamefont{Mueller}},
  \bibinfo{author}{\bibfnamefont{T.~L.} \bibnamefont{Ho}},
  \bibinfo{author}{\bibfnamefont{M.}~\bibnamefont{Ueda}}, \bibnamefont{and}
  \bibinfo{author}{\bibfnamefont{G.}~\bibnamefont{Baym}},
  \bibinfo{journal}{Phys. Rev. A} \textbf{\bibinfo{volume}{74}},
  \bibinfo{pages}{033612} (\bibinfo{year}{2006}).

\bibitem[{\citenamefont{Chen et~al.}(2007)\citenamefont{Chen, Chien, He, and
  Levin}}]{ChenStripes}
\bibinfo{author}{\bibfnamefont{Q.~J.} \bibnamefont{Chen}},
  \bibinfo{author}{\bibfnamefont{C.-C.} \bibnamefont{Chien}},
  \bibinfo{author}{\bibfnamefont{Y.}~\bibnamefont{He}}, \bibnamefont{and}
  \bibinfo{author}{\bibfnamefont{K.}~\bibnamefont{Levin}}, \bibinfo{journal}{J.
  Supercond. Nov. Magn.} \textbf{\bibinfo{volume}{20}}, \bibinfo{pages}{515}
  (\bibinfo{year}{2007}).

\bibitem[{\citenamefont{Zachar et~al.}(1997)\citenamefont{Zachar, Kivelson, and
  Emery}}]{Zachar1997}
\bibinfo{author}{\bibfnamefont{O.}~\bibnamefont{Zachar}},
  \bibinfo{author}{\bibfnamefont{S.~A.} \bibnamefont{Kivelson}},
  \bibnamefont{and} \bibinfo{author}{\bibfnamefont{V.~J.} \bibnamefont{Emery}},
  \bibinfo{journal}{J. Supercond.} \textbf{\bibinfo{volume}{10}},
  \bibinfo{pages}{373} (\bibinfo{year}{1997}).

\bibitem[{\citenamefont{Jaksch et~al.}(1998)\citenamefont{Jaksch, Bruder,
  Cirac, Gardiner, and Zoller}}]{Jaksch}
\bibinfo{author}{\bibfnamefont{D.}~\bibnamefont{Jaksch}},
  \bibinfo{author}{\bibfnamefont{C.}~\bibnamefont{Bruder}},
  \bibinfo{author}{\bibfnamefont{J.~I.} \bibnamefont{Cirac}},
  \bibinfo{author}{\bibfnamefont{C.~W.} \bibnamefont{Gardiner}},
  \bibnamefont{and} \bibinfo{author}{\bibfnamefont{P.}~\bibnamefont{Zoller}},
  \bibinfo{journal}{Phys. Rev. Lett.} \textbf{\bibinfo{volume}{81}},
  \bibinfo{pages}{3108} (\bibinfo{year}{1998}).

\bibitem[{\citenamefont{Petrov et~al.}(2004)\citenamefont{Petrov, Salomon, and
  Shlyapnikov}}]{Petrov}
\bibinfo{author}{\bibfnamefont{D.~S.} \bibnamefont{Petrov}},
  \bibinfo{author}{\bibfnamefont{C.}~\bibnamefont{Salomon}}, \bibnamefont{and}
  \bibinfo{author}{\bibfnamefont{G.~V.} \bibnamefont{Shlyapnikov}},
  \bibinfo{journal}{Phys. Rev. Lett.} \textbf{\bibinfo{volume}{93}},
  \bibinfo{pages}{090404} (\bibinfo{year}{2004}).

\end{thebibliography}

\end{document}